\documentclass[11pt, letterpaper]{article}

\usepackage{ifthen}
\newcommand{\Bset}[1]{\setboolean{#1}{true}}
\newcommand{\Bunset}[1]{\setboolean{#1}{false}}

\usepackage{amssymb}
\usepackage{amsmath}   %
\usepackage{amsthm}   %
\usepackage{latexsym}
\usepackage{amsfonts}
\usepackage{psfrag}    %
\usepackage{url}    %

\usepackage[dvips]{graphicx}
\usepackage[usenames,dvipsnames]{color}

\newcommand{\Bif}[3]{\ifthenelse{\boolean{#1}}{#2}{#3}}

{\theoremstyle{remark}}

{\theoremstyle{definition}}

\newtheoremstyle{anystatement}{\topsep}{\topsep}{\itshape}{}{\bfseries}{.}{ }{\anystatementname}
{\theoremstyle{anystatement}}
\newcommand{\anystatementname}{}

\newcounter{tmp_id_cnt}
\newcommand{\nospell}[1]{#1}   %

\newcommand{\mydef}[2]{\def#1{#2}}

\newcommand{\newident}[3][*]{\ifthenelse{\equal{*}{#1}}  %
{\newcommand{#2}[1][*]                                 %
{\ifthenelse{\equal{*}{##1}}                          %
{\nospell{\mbox{\Ensuremath{{\mathit{#3}}}}}}        %
{\ifthenelse{\equal{b}{##1}}                         %
{\nospell{\mbox{\Ensuremath{{\mathbf{#3}}}}}}      %
{#3}}}}                                            %
{\mydef{#2}{#3}}}                                      %

\newcommand{\newmat}[3][*]{\ifthenelse{\equal{*}{#1}}              %
{\newcommand{#2}[1][*]                                           %
{\ifthenelse{\equal{b}{##1}}                                    %
{\nospell{\mbox{\Ensuremath{\mathbf{#3}}}}}                    %
{\ifthenelse{                                                  %
\( \equal{*}{##1} \and \not \boolean{in_math_mode} \)         %
\or \( \not \equal{*}{##1} \and \boolean{in_math_mode} \)}    %
{\nospell{\mbox{\Ensuremath{#3}}}}                            %
{#3}}}}                                                       %
{\mydef{#2}{#3}}}                                                %

\newcommand{\newmatarg}[2]{                                              %
\newcommand{#1}[1][]                                                   %
{\ifthenelse{\boolean{in_math_mode}}                                  %
{#2}                                                                %
{\nospell{\mbox{\Ensuremath{#2}}}}}}                                %

\newcommand{\newmatop}[2]{\mydef{#1}{\operatorname{#2}}}

\newcommand{\MyMakeTheoMacros}[3]{
\newcommand{#2}[2][]{\ifthenelse{\equal{}{##1}}
{\begin{#1} ##2 \end{#1}}
{\begin{#1}\label{##1} ##2\end{#1}}}
\newcommand{#3}[3][]{\ifthenelse{\equal{}{##1}}
{\begin{#1}{\e{##2}} ##3 \end{#1}}
{\begin{#1}{\e{##2}}\label{##1} ##3\end{#1}}}
}

\newcommand{\MyMakeDupTheoMacros}[8]{
\MyMakeTheoMacros{#1}{#2}{#3}
\newcommand{#4}[3]{
\newcommand{##2}{##3}
\begin{#1}\label{##1} ##2\end{#1}}
\newcommand{#5}[4]{
\newcommand{##2}{##4}
\begin{#1}{\e{##3}}\label{##1} ##2\end{#1}}
\newcommand{#8}[2]{\def\my_tmp_id{my_tmp_id_\arabic{tmp_id_cnt}}
\newtheorem*{\my_tmp_id}{#7~\ref{##1}}
\begin{\my_tmp_id} ##2 \end{\my_tmp_id}\stepcounter{tmp_id_cnt}}
\newcommand{#6}[6]{
#2[##1]{##2}

##3
\prf[#7~\ref{##1}]{##6} \newcommand{##5}{}

}
}

\newcommand{\MyMakeRefMacros}[3]{\newcommand{#1}[2][]
{\ifthenelse{\equal{}{##1}}{#2~\ref{##2}}{#3~\ref{##1} and~\ref{##2}}}}

\newcommand{\MyMakeEqRefMacros}[3]{\newcommand{#1}[2][]
{\ifthenelse{\equal{}{##1}}{#2~\eqref{##2}}{#3~\eqref{##1} and~\eqref{##2}}}}

\newcommand{\abstr}[1]{
\begin{abstract}
#1
\end{abstract}}

\newcommand{\bibentry}[8]{

\bibitem[\nospell{#8}]{#1} {\textup #3}. 
\ifthenelse{\equal{}{#6}}
{\newblock \textrm{#4.} \newblock {\em #5}, #7.}
{\newblock \textrm{#4.} \newblock {\em #5, #6}, #7.}
}

\newcommand{\inputbib}{

\bibentryPerCom[2006]{H. Buhrman}{B}

\bibentry{BCW98}{Buhrman, Cleve and Wigderson}{H. Buhrman, R. Cleve and A. Wigderson}{Quantum vs. Classical Communication and Computation}{Proceedings of the 30th Symposium on Theory of Computing}{pages 63-68}{1998}{BCW98}

\bibentry{BCWW01}{Buhrman, Cleve, Watrous and de Wolf}{H. Buhrman, R. Cleve, J. Watrous and R. de Wolf}{Quantum Fingerprinting}{Physical Review Letters 87(16)}{article 167902}{2001}{BCWW01}

\bibentry{BFS86}{Babai, Frankl and Simon}{L. Babai, P. Frankl and J. Simon}{Complexity classes in communication complexity theory}{Proceedings of the 27th Annual Symposium on Foundations of Computer Science}{pages 337-347}{1986}{BFS86}

\bibentry{BJK04}{Bar-Yossef, Jayram and Kerenidis}{Z. Bar-Yossef, T. S. Jayram and I. Kerenidis}{Exponential Separation of Quantum and Classical One-Way Communication Complexity}{Proceedings of 36th Symposium on Theory of Computing}{pages 128-137}{2004}{BJK04}

\bibentryPerCom[2005]{R. Cleve}{C}

\bibentry{GKKRW07}{Gavinsky, Kempe, Kerenidis, Raz and de Wolf}{D. Gavinsky, J. Kempe, I. Kerenidis, R. Raz and R. de Wolf}{Exponential Separations for One-Way Quantum Communication Complexity, with Applications to Cryptography}{Proceedings of the 39th Symposium on Theory of Computing}{pages 516-525}{2007}{GKKRW07}

\bibentry{JKN08_Dir}{Jain, Klauck and Nayak}{R. Jain, H. Klauck and A. Nayak}{Direct Product Theorems for Classical Communication Complexity via Subdistribution Bounds}{Proceedings of the 40th Symposium on Theory of Computing}{}{2008}{JKN08}

\bibentry{KN97}{Kushilevitz and Nisan}{E. Kushilevitz and N. Nisan}{Communication Complexity}{Cambridge University Press}{}{1997}{KN97}

\bibentry{R92}{Razborov}{A. Razborov}{On the Distributional Complexity of Disjointness}{Theoretical Computer Science 106(2)}{pages 385-390}{1992}{R92}

\bibentry{R99}{Raz}{R. Raz}{Exponential Separation of Quantum and Classical Communication Complexity}{Proceedings of the 31st Symposium on Theory of Computing}{pages 358-367}{1999}{R99}

\bibentry{Y79}{Yao}{A. C-C. Yao}{Some Complexity Questions Related to Distributed Computing}{Proceedings of the 11th Symposium on Theory of Computing}{pages 209-213}{1979}{Y79}

}

\newcommand{\bib}[1][]{

\begin{thebibliography}{ABCDEFG00}
\frenchspacing
\inputbib
#1
\end{thebibliography}
}

\newcommand{\citePerCom}[2][]{\cite{#2}}
\newcommand{\bibentryPerCom}[3][]{ \ifthenelse{\equal{}{#1}}
{
\bibitem[\nospell{#3}]{#2} {\textup #2} - \textit{Personal communication}. }
{
\bibitem[\nospell{#3}]{#2} {\textup #2} - \textit{Personal communication}, #1. 
}
}

\MyMakeTheoMacros{fact}{\fct}{\nfct}

\MyMakeRefMacros{\fctref}{Fact}{Facts}

\MyMakeDupTheoMacros{lemma}
{\lem}{\nlem}{\lemdup}{\nlemdup}{\lemapp}{Lemma}{\lemrep}

\MyMakeRefMacros{\lemref}{Lemma}{Lemmas}

\MyMakeDupTheoMacros{corollary}
{\crl}{\ncrl}{\crldup}{\ncrldup}{\crlapp}{Corollary}{\crlrep}

\MyMakeRefMacros{\crlref}{Corollary}{Corollaries}

\MyMakeTheoMacros{proposition}{\pr}{\npr}

\newtheorem*{prp*}{\e{Proposition}}
\newcommand{\prpunnambered}[1]{\begin{prp*} #1 \end{prp*}}
\mydef{\prp*}{\prpunnambered}

\MyMakeRefMacros{\prpref}{Proposition}{Propositions}

\MyMakeDupTheoMacros{claim}
{\clm}{\nclm}{\clmdup}{\nclmdup}{\clmapp}{Claim}{\clmrep}

\MyMakeRefMacros{\clmref}{Claim}{Claims}

\MyMakeDupTheoMacros{theorem}
{\theo}{\ntheo}{\theodup}{\ntheodup}{\theoapp}{Theorem}{\theorep}

\MyMakeRefMacros{\theoref}{Theorem}{Theorems}

\MyMakeTheoMacros{definition}{\defi}{\ndefi}

\MyMakeRefMacros{\defiref}{Definition}{Definitions}

\MyMakeTheoMacros{problem}{\prob}{\nprob}

\MyMakeRefMacros{\probref}{Problem}{Problems}

\MyMakeTheoMacros{protocol}{\prot}{\nprot}

\MyMakeRefMacros{\protref}{Protocol}{Protocols}

\providecommand{\qedsymbol}{\square}

\newcommand{\prf}[2][]{\ifthenelse{\equal{}{#1}}
{\begin{proof}\renewcommand{\qedsymbol}{$\blacksquare$} #2 \end{proof}}
{\begin{proof}[Proof of #1]
\renewcommand{\qedsymbol}{$\blacksquare_{\mbox{\it{\scriptsize{#1}}}}$}
#2 \end{proof}}
}

\newcommand{\sect}[2][]{\ifthenelse{\equal{}{#1}}
{\section{#2}}
{\section{#2}\label{#1}}}

\newcommand{\ssect}[2][]{\ifthenelse{\equal{}{#1}}
{\subsection{#2}}
{\subsection{#2}\label{#1}}}

\MyMakeRefMacros{\chref}{Chapter}{Chapters}

\MyMakeRefMacros{\sref}{Section}{Sections}

\MyMakeRefMacros{\ssref}{Subsection}{Subsections}

\MyMakeRefMacros{\sssref}{Subsection}{Subsections}

\definecolor{DarkGreen}{rgb}{0,0.45,0.08}

\newboolean{in_math_mode}
\setboolean{in_math_mode}{false}

\newcommand{\IfMathMode}{\Bif{in_math_mode}}

\newcommand{\MathModeOn}{\Bset{in_math_mode}}

\newcommand{\MathModeOff}{\Bunset{in_math_mode}}

\newcommand{\Ensuremath}[1]{\IfMathMode   %
{\ensuremath{#1}}
{\MathModeOn\ensuremath{#1}\MathModeOff}}

\newcommand{\fbr}[1]{\IfMathMode   %
{$#1$}                              %
{\MathModeOn$#1$\MathModeOff}}

\newcommand{\fnbr}[1]{\mbox{\fbr{#1}}}   %

\newcommand{\fla}[2][*]{\ifthenelse{\equal{}{#1}}{\fbr{#2}}{\fnbr{#2}}}

\newcommand{\bfla}[2][]{\mbox{\fbr{\mathbf{#2}}}}

\newcommand{\mat}[2][]{\ifthenelse{\equal{}{#1}}   %
{\begin{displaymath} \MathModeOn
#2
\MathModeOff \end{displaymath}     %
}{
\begin{equation} \MathModeOn \label{#1}
#2
\MathModeOff \end{equation}
}
}

\newcommand{\matal}[2][]{\mat[#1]{\begin{split} #2 \end{split}}}

\newcommand{\f}{\fla}

\newcommand{\fb}{\bfla}

\newcommand{\m}{\mat}

\newcommand{\mal}{\matal}

\newcommand{\mac}{\substack}    %

\newcommand{\twocase}[4]
{\begin{cases}#1 &\txt{#2}\\#3 &\txt{#4}\end{cases}}

\MyMakeEqRefMacros{\equref}{Equation}{Equations}

\MyMakeEqRefMacros{\expref}{Expression}{Expressions}

\MyMakeEqRefMacros{\inequref}{Inequality}{Inequalities}

\newcommand{\bracref}[1]{(\ref{#1})}

\newcommand{\bref}{\bracref}

\MyMakeRefMacros{\figref}{Figure}{Figures}

\providecommand{\middle}{\big}

\newcommand{\chs}{\genfrac(){0cm}{}}   %

\newmatop{\mymod}{mod}   %
\newmatop{\poly}{poly}
\newmatop{\sign}{sign}
\newmatop{\tr}{tr}
\newmatop{\supp}{supp}    %
\newmatop{\argmax}{argmax}
\newmatop{\argmin}{argmin}
\newmatop{\posmin}{posmin}
\newmatop{\negmin}{negmin}
\newmatop{\posmax}{posmax}
\newmatop{\negmax}{negmax}
\newmatop{\argmaxmin}{argmax(min)}
\newcommand{\smcup}{\mathop{\cup}}

\newcommand{\h}[2][]{\ifthenelse{\equal{}{#2}}
{\mathop{\mathbf{H}}_{#1}}
{\mathop{\mathbf{H}}_{#1}\left[{#2}\right]}}
\newcommand{\hh}[3][]{\mathop{\mathbf{H}}_{#1}\left[{#2}\middle|\vphantom{|_1^1}{#3}\right]}
\newcommand{\E}[2][]{\mathop{\mathbf{E}}_{#1}\left[{#2}\right]}

\newcommand{\PR}[2][]{\mathop{\mathbf{Pr}}_{#1}\left[{#2}\right]}
\newcommand{\PRr}[3][]{\mathop{\mathbf{Pr}}_{#1}\left[{#2}\middle|\vphantom{|_1^1}{#3}\right]}

\newcommand{\U}[1][]{\ifthenelse{\equal{}{#1}}
{{\cal U}}
{{\cal U}_{#1}}}

\newcommand{\Uu}[2]{\ifthenelse{\equal{}{#1}}
{{\cal U}^{#2}}
{{\cal U}_{#1}^{(#2)}}}

\newcommand{\GF}[1]{{\cal GF}_{#1}}

\newcommand{\ov}{\overline}

\newcommand{\pl}[1][]{\nospell{\ifthenelse{\equal{}{#1}}
{\mbox{-s}}
{\fla{#1}\mbox{-s}}}}

\newmat[]{\llp}{\left(}
\newmat[]{\rrp}{\right)}

\newmat[]{\lla}{\left\langle}
\newmat[]{\rra}{\right\rangle}

\newmat[]{\llf}{\left\lfloor}
\newmat[]{\rrf}{\right\rfloor}

\newmat[]{\dt}{\cdot}
\newmat[]{\tm}{\cdot}
\newmat[]{\sbseq}{\subseteq}
\newmat[]{\sbs}{\subset}
\newmat[]{\smin}{\setminus}
\newmat[]{\mset}{\smin\set}
\newmat[]{\eps}{\varepsilon}
\newmat[]{\deq}{\stackrel{\textrm{def}}{=}}

\newmat{\NN}{\mathbb{N}}

\newmat[]{\ds}{\dots}
\newmat[]{\dc}{,\ds,}
\newmat[]{\dcup}{\cup\ds\cup}

\newcommand{\enum}[1]{\begin{enumerate} #1 \end{enumerate}}

\newcommand{\descr}[1]{\begin{description} #1 \end{description}}

\newcommand{\itemi}[2][]{\ifthenelse{\equal{}{#1}}
{\begin{itemize} #2 \end{itemize}}
{\begin{itemize}[#1] #2 \end{itemize}}}

\newcommand{\wrt}{w.r.t.\ }	 %

\newcommand{\ie}{i.e., }	 %

\newcommand{\eg}{e.g., }	 %

\newcommand{\st}{such that\ }		 %

\newcommand{\fr}[3][*]{
\ifthenelse{\equal{*}{#1}}               %
{\frac{#2}{#3}}{}
\ifthenelse{\equal{}{#1}}                %
{\left.#2\middle/#3\right.}{}
\ifthenelse{\equal{b_}{#1}}              %
{\left.\left(#2\right)\middle/#3\right.}{}
\ifthenelse{\equal{_b}{#1}}              %
{\left.#2\middle/\left(#3\right)\right.}{}
\ifthenelse{\equal{bb}{#1}}              %
{\left.\left(#2\right)\middle/\left(#3\right)\right.}{}
}

\newcommand{\set}[2][]{\ifthenelse{\equal{}{#1}}               %
{\Ensuremath{\left\{#2\right\}}}                             %
{\Ensuremath{\left\{#2\middle|\vphantom{|_1^1}#1\right\}}}}  %

\newcommand{\Max}[2][]{\ifthenelse{\equal{}{#1}}   %
{\Ensuremath{\max\left\{#2\right\}}}             %
{\Ensuremath{\max_{#1}\left\{#2\right\}}}}       %

\newcommand{\newfunction}[2]{                          %
\newcommand{#1}[2][*]{\ifthenelse{\equal{*}{##1}}    %
{\Ensuremath{#2\left(##2\right)}}                   %
{#2(##2)}}                                          %
}                                                     %

\newfunction{\asO}{O}
\newfunction{\astO}{\tilde O}
\newfunction{\aso}{o}
\newfunction{\asOm}{\Omega}
\newfunction{\astOm}{\tilde \Omega}
\newfunction{\asom}{\omega}
\newfunction{\asT}{\Theta}

\newcommand{\ket}[1]{\Ensuremath{\left|#1\rra}}
\newcommand{\kbra}[2][]{\ifthenelse{\equal{}{#1}}
{\Ensuremath{\left|#2\rra\hspace{-3.5pt}\lla #2\right|}}
{\Ensuremath{\left|#1\rra\hspace{-3.5pt}\lla #2\right|}}
}

\newmat[]{\01}{\set{0,1}}

\newcommand{\sz}[2][]{\ifthenelse{\equal{}{#1}}
{\Ensuremath{\left|#2\right|}}
{\Ensuremath{\left|#2\right|_{#1}}}}

\newcommand{\lra}[2][*]{\ifthenelse{\equal{}{#1}}
{\langle #2 \rangle}
{\lla #2 \rra}}
\newcommand{\floor}[2][*]{\ifthenelse{\equal{}{#1}}
{\lfloor #2 \rfloor}
{\llf #2 \rrf}}

\newcommand{\fn}{\footnote}

\newcommand{\nin}{\not\in}   %

\newcommand{\e}{\emph}

\newcommand{\il}[1]{{\it #1}}  %

\newfont{\bit}{cmbxsl10}

\newcommand{\txt}[1]{\textrm{#1}}   %

\newcommand{\tit}[1]{\textit{#1}}   %

\newcommand{\tbb}{\qquad}

\newident{\R}{\mathcal R}
\newident{\RI}{\mathcal R^{1}}
\newident{\Q}{\mathcal Q}
\newident{\QI}{\mathcal Q^{1}}

\usepackage{hyphenat}   %

\newcommand{\whereproofs}{Appendix} 

\date{}
\newmatarg{\Uii}{\ifthenelse{\equal{}{#1}}
{\Uu{1\times 1}{n}}
{\Uu{1\times 1}{n;#1}}}
\newmatarg{\UA}{\ifthenelse{\equal{}{#1}}
{\U[A]}
{\Uu{A}{#1}}}
\newmat{\Ua}{\U[\A]}
\newmat{\Ub}{\U[\B]}
\newmatarg{\UAa}{\ifthenelse{\equal{}{#1}}
{\Uu{A}{\!\A}}
{\Uu{A}{\!\A;#1}}}
\newmatarg{\UAb}{\ifthenelse{\equal{}{#1}}
{\Uu{A}{\!\B}}
{\Uu{A}{\!\B;#1}}}
\newmatarg{\UB}{\ifthenelse{\equal{}{#1}}
{\U[B]}
{\Uu{B}{#1}}}

\newmatarg{\pt}{\ifthenelse{\equal{}{#1}}
{p_t^{(t)}}
{p_{#1}^{(t)}}}

\newident{\Pin}{P}

\newident{\PS}{P_{1\times 1}^{\Sigma}}

\newident{\Piip}{P_{1\times 1}^{\txt{\tiny \it search}}}
\newident{\X}{{\cal X}_1}
\newident{\cR}{\bf R}
\newident{\cX}{\bf X}
\newident{\cY}{\bf Y}
\newident{\Xo}{{\cal X}_0}
\newident{\Xii}{{\cal X}_2}
\newident{\Xj}{{\cal X}_j}
\newident{\Ec}{{\cal E}}
\newident{\D}{DISJ}
\newident{\Dn}{DISJ_n}
\newident{\Dm}{DISJ_m}

\newcommand{\A}{\tit{Alice}}
\newcommand{\sA}{\|_{\A}}
\newcommand{\B}{\tit{Bob}}
\newcommand{\sB}{\|_{\B}}

\title{Classical Interaction Cannot Replace a Quantum Message}

\author{
{\bf Dmitry Gavinsky}
\thanks{Part of this work was done while at the Institute for Quantum Computing at the University of Waterloo.} \\
{\small NEC Laboratories America, Inc.}\\
{\small 4 Independence Way, Suite 200}\\
{\small Princeton, NJ 08540, U.S.A.}
}

\begin{document}

\maketitle

\thispagestyle{empty}

\abstr{We demonstrate a two-player communication problem that can be solved in the one-way quantum model by a \f0-error protocol of cost \asO{\log n} but requires exponentially more communication in the classical interactive (bounded error) model.}

\setcounter{page}{1}

\sect{Introduction}
The ultimate goal of quantum computing is to identify computational tasks that by using the laws of quantum mechanics can be solved more efficiently than on a classical computer.

In this paper we study quantum computation from the perspective of Communication Complexity, first defined by Yao~\cite{Y79}.
Two parties, Alice and Bob, try to solve a computational problem that depends on $x$ and $y$.
Initially Alice knows only $x$ and Bob knows only $y$; in order to solve the problem they communicate, obeying to restrictions of a specific \e{communication model}.
In order to compare the power of two communication models, one has to either prove existence of a task that can be solved more efficiently in one model than in the other, or to argue that no such task exists.

We will, in the first place, be concerned about the following models.\itemi{
\item \e{One-way communication} is the model where Alice sends a single message to Bob who has to give an answer, based on the content of the message and his part of input.
\item \e{Interactive (two-way) communication} is the model where the players can interactively exchange messages till Bob decides to give an answer, based on the communication transcript and his part of input.
}
Both models can be either \e{classical} or \e{quantum}, according to the nature of communication allowed between the players.
The classical versions of the models are denoted by \RI\ and \R, and the quantum versions are denoted by \QI\ and \Q, respectively.
It is clear that interactive communication is at least as powerful as one-way communication, and it is well-known that the former can sometimes be much more efficient than the latter, both in quantum and in classical versions.

Communication tasks can be either \e{functional}, meaning that there is exactly one correct answer corresponding to every possible input, or \e{relational}, when multiple correct answers are allowed.
Functional tasks over domains forming product sets \wrt each players' inputs are called \e{total}.

A \e{communication protocol} describes behavior of Alice and Bob in response to each possible input.
The \e{cost} of a protocol is the maximum total amount of (qu)bits communicated by the parties, according to the protocol.

We say that a communication task $P$ is solvable \e{with bounded error} in a given communication model by a protocol of cost \asO{k} if for any constant $\eps>0$, there exists a corresponding protocol solving $P$ with success probability at least $1-\eps$.
If the protocols, in addition, either refuse to answer or succeed, then we say that the solution is \e{\f0-error}.

In this paper our primary concern is with separating communication models; more specifically, with finding communication problems that demonstrate super-polynomial advantage of quantum communication over classical one.
In fact, both with the previously known examples considered below and with our own contribution it has been the case that the first shown super-polynomial separation had actually been exponential.

It is important to note that the three types of communication tasks mentioned above (relational, functional and total functional) form a \e{hierarchy}, if viewed as \e{tools to separate communication model}.
In particular, there are known pairs of communication models that can be separated through a relational problem but are equally strong over functions, either total or partial; likewise, there are pairs of communication models that can be separated through a partial functional problem but are \e{widely conjectured} to be equally strong over total functions.

For \f0-error, both one-way and interactive protocols, separations have been demonstrated by Buhrman, Cleve, and Wigderson~\cite{BCW98}.
In the bounded-error setting the first separation has been given by Raz~\cite{R99}, showing a problem solvable in \Q\ exponentially more efficiently than in \R.
Later, Buhrman, Cleve, Watrous, and de Wolf~\cite{BCWW01} demonstrated an exponential separation for \e{simultaneous protocols}, which is a communication model even more limited than one-way.
All these separations have been given for functional problems.

For one-way protocols with bounded error, the first separation has been shown by Bar-Yossef, Jayram, and Kerenidis~\cite{BJK04} for a relational problem.
Later, Gavinsky, Kempe, Kerenidis, Raz, and de Wolf~\cite{GKKRW07} gave a similar separation for a partial functional problem.

These results show that quantum communication models can be very efficient, when compared to their classical counterparts.
But \e{does there exist a problem that can be solved by a quantum one-way protocol more efficiently than by any classical two-way protocol?}

\ssect{Our result}
\theo[theo_main]{For infinitely many $N\in\NN$, there exists an (explicit) relation with input length $N$ that can be solved by a \f0-error one-way quantum protocol of cost \asO{\log N} and whose complexity in the interactive classical model is \asOm{\fr{N^{\fr[]18}}{\sqrt{\log N}}}.}

This statement simultaneously subsumes the separation in~\cite{BJK04} and, as our theorem speaks about a relational problem, partially that in~\cite{R99}.
To obtain a similar result for a functional problem is an important open question (see \sref{s_open} for more).

The relation we use is a modification of a communication task independently suggested by R.\ Cleve (\citePerCom[2005]{R. Cleve}) and S.\ Massar (\citePerCom[2006]{H. Buhrman}) as a possible candidate for such separation.

Some of the intermediate steps in our proof might be of independent interest.

\sect[s_appr]{Our approach}
Denote by~$\bar0$ the additive identity of a field.
For $n$ being a power of $2$, define the following communication problems.

\defi{Let $x, y\sbs [n^2]$, such that $|x|=n/2$ and $|y|=n$.
Let $z\in\GF2^{2\log n}\mset{\bar0}$.
Let  $\Sigma=\set{\sigma_{2^{2i}}}_{i=1}^\infty$ be a set of reversible mappings from $[2^{2i}]$ to $\GF2^{2i}$.
Then $(x,y,z)\in \PS$ if either $|x\cap y|\neq2$ or $\lra{z,a+b}=0$, where $\sigma_{n^2}(x\cap y)=\set{a,b}$.}

Let $\Sigma_0$ be the set of reversible mappings from $[2^{2i}]$ to $\GF2^{2i}$, preserving the lexicographic ordering of the elements.

\defi{Let $x\sbs [2n^2]$, $|x|=n$.
Let $y=(y_1\dc y_{n/4})$ be a tuple of disjoint subsets of $[n^2]$, each of size $n$, such that $|x\cap y_j|=2$ for all $1\le j\le n/4$.
Let $z\in\GF2^{2\log n+1}\mset{\bar0}$ and $1\le i\le n/4$, then $(x,y,(i,z))\in P^{(n)}$ if $\lra{z,a+b}=0$, where $\sigma_0(x\cap y_i)=\set{a,b}$ for some $\sigma_0\in\Sigma_0$.}

In the rest of the paper we will implicitly assume equivalence between the arguments and the corresponding values of every $\sigma_0\in\Sigma_0$.

We will show that \Pin\ is easy to solve in \QI\ and is hard for \R.
In order to prove the lower bound we will use the following modification of \PS.

\defi{Let $x, y\sbs [n^2]$, such that $|x|=n/2$ and $|y|=n$.
Let $z\sbs [n^2]$.
Then $(x,y,z)\in \Piip$ if $x\cap y=z.$}

We consider correctness probability of communication protocols \wrt input distribution, or the randomness of the protocol, or both.
Unless stated otherwise, all available randomness is taken into account, \ie input distribution is considered whenever it is known and protocol's randomness is considered unless the protocol under consideration is deterministic.

We use the following generalization of the standard bounded error setting.
We say that a protocol solves a problem \e{with probability $\delta$ with error bounded by $\eps$} if with probability at least $\delta$ the protocol produces an answer, and whenever produced, the answer is correct with probability at least $1-\eps$.

Solving \PS\ when $|x\cap y|=2$ requires providing an evidence of knowledge of these elements, and intuitively should be as hard as finding them, as required by \Piip, when $|x\cap y|=2$.
This intuition is most likely \e{false} for the quantum 1-way model (when $|x\cap y|=2$, \PS\ can be efficiently\fn
{In the context of communication complexity, \e{efficient} protocols are those of polylogarithmic cost.}
solved in \QI\ with probability $1/n$ with small error, which is unlikely to be the case for \Piip).
However, it is true for the model of classical 2-way communication; a ``quasi-reduction'' from \Piip\ to \PS\ is one of the central ingredients of our lower bound proof.

The high-level structure of the proof is the following.
\descr{
\item[Solution to {\Pin[b]} $\implies$ Solution to {\PS[b]} \il{(\lemref{l_in2ii})}] We claim that if there exists a protocol that solves \Pin\ with error bounded by $\eps$ then another protocol of similar cost solves \PS\ for some $\Sigma$ with probability \asOm{1/n} and error \asO{\eps}.
\item[Solution to {\PS[b]} $\implies$ Solution to {\Piip[b]} \il{(\theoref{t_ii2iip})}] We reduce the task of solving the problem \Piip\ to that of solving \PS.
\item[{\Piip[b]} is hard \il{(\theoref{t_hard_iip})}] We show that the cost of solving \Piip\ with probability $\delta$ when $\sz{x\cap y}=2$ is \asOm{n\dt\sqrt{\delta}}.
}
We will conclude that solving \Pin\ with bounded error requires an interactive classical protocol of complexity $n^{\asOm1}$.

\sect{Notation and more}
We assume basic knowledge of (classical) communication complexity~(\cite{KN97}).

We will consider only discrete probability distributions.
For a set $A$ we write $\U[A]$ to denote the uniform distribution over the elements of $A$.
Given a distribution $D$ over $A$ and some $a_0\in A$ we denote $D(a_0)\deq\PR[\cX\sim D]{\cX=a_0}$; for $B\sbseq A$, $D(B)\deq\sum_{b\in B}D(b)$.
Denote $\supp(D)\deq\set[D(a)>0]{a\in A}$.

Let $\cX$, $\cY$ be (discrete) random variables.
We let $\h{\cX}$ and $\hh{\cX}{\cY}$ denote the corresponding entropy and conditional entropy.
As a function of $\cY$, we will denote the conditional entropy by $\hh{\cX}{\cY=y}$.

We will need the Chernoff bound in the following form.
\clm[c_Cher]{Let $\cX_1\dc\cX_m$ be random variables, distributed independently and satisfying for some $\mu,\alpha>0$
\m{\forall 1\le i\le m:\:0\le\cX_i\le\alpha,\,\E{\cX_i}\le\mu~.}
Then
\m{\PR{\fr1m\tm\sum_{i=1}^m\cX_i\ge\llp1+\asOm1\rrp\tm\mu}
\in2^{-\asOm{\fr{m\mu}{\alpha}}}.}}

We use the following notation.
\mal{
\Dn&\deq\set[x,y\in\01^n,\,\forall i\in{[n]}:\:x_i=0\vee y_i=0]{(x,y)};\\
\D&\deq\cup_{n\in\NN}\Dn.
}

We use the standard notion of a \e{(combinatorial) rectangle}.
The sides of considered rectangle always correspond to subsets of the input sets of Alice and Bob, as defined by the communication problem under consideration (to emphasize this, we will sometimes use the term \e{input rectangle}).
We will use the same notation for an input rectangle and for the \e{event that the input belongs to the rectangle}.

Define context-sensitive ``projection operators'' $\dt|_\dt$ and $\dt\|_\dt$ as follows.
For a discrete set $A$, $x\sbseq A$ and $I\sbseq A$, let  $x|_I\deq x\cap I$.
For $B\sbseq2^A$, let \f{B\|_I\deq\set[x\in B]{x\cap I}}.
For a distribution $D$ over $A$, let $D|_I$ be the conditional distribution of $\cX\sim D$, subject to $\cX\in I$.
For a distribution $D$ over $2^A$, let $D\|_I$ be the marginal distribution of $\cY\deq \cX|_I$, when $\cX\sim D$.

We will use special notation for ``one-sided'' projections of input pairs.
Let $(x,y)\in {\cal A}\times{\cal B}$, where $\cal A$ and $\cal B$ are input sets of Alice and Bob, respectively.
Then $(x,y)|_{\A}\deq x$ and $(x,y)|_{\B}\deq y$.
Similarly, define the operators $\|_{\A}$ and $\|_{\B}$ for distributions and sets.

\ssect{More details on \PS\ and \Pin}
Define the following events characterizing input to \PS\ or \Piip.
\defi{For $j\in\NN$, let $\Xj$ be the event that the input pair $(x,y)$ satisfies $\sz{x\cap y}=j$.
For $i,j\in\NN$, let $\X(i)$ and $\Xii(i,j)$ be, respectively, the events that $x\cap y=\set i$ and $x\cap y=\set{i,j}$.}
We will use the same notation to address the subsets of input that give rise to these events, \ie 
\m{\Xo\deq\smcup_{n=2^i}\set[x\cap y=\emptyset]{(x,y)\in [n^2]\times [n^2]},}
and so forth.

We define \Uii\ to be the uniform distribution of input to \PS, $\Ua\deq\Uii\sA$ and $\Ub\deq\Uii\sB$.
\defi{For $k_1\dc k_t\in\NN$, let
\mal{
\Uii[k_1\dc k_t]&\deq\Uii|_{{\cal X}_{k_1}\dcup{\cal X}_{k_t}},\\
\Uii[\ge k_1]&\deq\Uii|_{\smcup_{i\ge k_1}{\cal X}_i}.}
}
\defi{Given input set $A$ (not necessarily a rectangle), define
\mal{
\UA&\deq\Uii|_A,\\
\UAa&\deq\UA\sA,\\
\UAb&\deq\UA\sB.}
Given $k_1\dc k_t\in\NN$, let
\m{\UA[k_1\dc k_t]\deq\UA|_{{\cal X}_{k_1}\dcup{\cal X}_{k_t}}.}}

\clm[c_X]{For sufficiently large $n$ it holds that $\Uii(\Xo)\ge1/3$, $\Uii(\X)\ge1/6$ and $\Uii(\Xii)\ge1/13$.
On the other hand, for any $t\le n/2$ it holds that $\Uii\llp\smcup_{i\ge t}{\cal X}_i\rrp\le\llp\fr34\rrp^t$.}

\prf[\clmref{c_X}]{Think about choosing input pair $(\cX,\cY)\sim\Uii$ as selecting a random subset $\cY\sbs [n^2]$, subject to $|\cY|=n$, followed by selecting $n/2$ distinct elements for $\cX$.
Under such interpretation it is clear that $\Uii\llp\smcup_{i\ge t}{\cal X}_i\rrp\le{\chs{n/2}t}\tm\llp\fr{n}{n^2-n/2}\rrp^t$.
Therefore, $\Uii(\Xo)\ge1-n/2\tm\fr{n}{n^2-n/2}\ge\fr13$ and $\Uii\llp\smcup_{i\ge t}{\cal X}_i\rrp\le\llp\fr n2\rrp^t\tm\llp\fr3{2n}\rrp^t=\llp\fr34\rrp^t$, for $n\ge2$.

Let $E_i$ be the event that $i\in\cX\cap\cY$.
It clearly follows from the symmetry between all \pl[E_i] and from the fact that the events are mutually exclusive when conditioned upon \X, that $\Uii(\X)$ is equal to $n/2$ times the probability that the first element selected for $\cX$ belongs to $\cY$ and all the following are not from $\cY$.
The former occurs with probability at least $1/n$ and the latter with probability not smaller than $\Uii(\Xo)$, therefore $\Uii(\X)\ge\fr n2\tm\fr1n\tm\fr13\ge\fr16$.

Similarly, $\Uii(\Xii)\ge{\chs{n/2}2}\tm\fr1n\tm\fr{n-1}{n^2-1}\tm\Uii(\Xo)>\fr1{13}$, for sufficiently large $n$.}

\ssect{Size of near-monotone rectangles for \Dn}
Two following lemmas can be viewed as the core of our lower bound proof, from both conceptual and technical points of view.

In his elegant lower bound proof for \D, Razborov~\cite{R92} has established the following lemma.
\nlem[razrazlem]{\cite{R92}}{Let $A$ be an input rectangle for \Dn, assume that $n=4l-1$.
Let $D$ be the following input distribution -- with probability $3/4$ Alice and Bob receive two uniformly distributed disjoint subsets of $[n]$ of size $l$ and with probability $1/4$ they receive two uniformly distributed subsets of $[n]$ of size $l$ that share exactly one element.
Then
\m{D(A\cap\X)\ge\fr1{135}\dt D(A\cap\Xo)-2^{-\asOm n}.}}

We need the following consequence of \lemref{razrazlem}.\fn
{Our \lemref{razlem} is similar to a statement made in the original lower bound proof of \asOm{\sqrt n} for \D\ by Babai, Frankl and Simon~\cite{BFS86}.
They consider a product distribution similar to our $D$ and give a lower bound on $D(A\setminus\Xo)$ in terms of $D(A\cap\Xo)$, while we need a lower bound on $D(A\cap\X)$.
We found extending the approach of \cite{BFS86} to be technically more challenging than deriving our statement from a stronger (non-product) case of \lemref{razrazlem}.}
\lemapp{razlem}
{Let $n$ be sufficiently large and $A$ be an input rectangle for \Dn.
Let $D$ be a product distribution of the two halves of the input, such that Alice receives a uniformly chosen subset of $[n]$ of size $k_1(n)$ and Bob receives a uniformly chosen subset of $[n]$ of size $k_2(n)$, where $\alpha_1\sqrt n\le k_1(n)\le k_2(n)\le\alpha_2\sqrt n$ for some $\alpha_1$, $\alpha_2$.
Then for $\delta=\fr{{\alpha_1}^2}{45\tm16^{{\alpha_2}^2}}$ it holds that
\m{D(A\cap\X)\ge\delta\tm D(A\cap\Xo)-2^{-\asOm{\sqrt n}}.}}
{}
{The proof can be found in the \whereproofs.}
{\apprazlem}
{We will reduce the communication task considered in \lemref{razrazlem} to that defined in the lemma we are proving.
Address the former task by $P'$ and the latter one by $P$ (they both are, in fact, versions of \D, defined \wrt different distributions).
We will use $m$ to denote the input length to $P'$.
The distribution of input to $P'$ corresponding to $m$ will be denoted by $D_m'$
The length and the distribution of input to $P$ will be denoted by $n$ and $D$, respectively.

Let $m=4k_1(n)-1$.
Let $T_r$ be a transformation $(x',y')\to(x,y)$, where $r\in\01^*$, $x',y'\in\01^{[m]}$, and $x,y\in\01^{[n]}$.
Think of $\cR=r$ as a uniform random string of sufficient length (we will address this situation by ``$\cR\sim\U$'') and of $T$ as a \e{randomized} transformation of $x'$ and $y'$ only (random bits are implicitly taken from $r$).
In order to compute $T_r(x',y')$ choose randomly and uniformly a pair $(M,\beta)$ of disjoint subsets of $[n]$ of sizes $m$ and $k_2(n)-l$, respectively (our choice of $n$ guarantees that the latter value is not negative).
Define $(x,y)$ by $x|_M=x'$, $y|_M=y'$, $x|_{\ov M}=\emptyset$ and $y|_{\ov M}=\beta$.
Note that $T$ can be applied locally by Alice and Bob if they share public randomness (that is, $x$ only depends on $r$ and $x'$ and $y$ only depends on $r$ and $y'$).

We can see that $(x,y)$ is input to \Dn\ and $\Dn(x,y)=\Dm(x',y')$, so indeed $T$ is a reduction from $\Dm$ to $\Dn$.
If $(x',y')$ comes from ${\cal X}_i\cap\supp(D_m')$ and $\cR\sim\U$ then $T_r(x',y')$ is uniformly distributed over ${\cal X}_i\cap\supp(D)$, for any $i\ge0$.
In particular, for $i\in\01$,
\m{\E[{\cR[]}\sim\U]{\PR[{(x',y')\sim{D_m'}|_{{\cal X}_i}}]{T_r(x',y')\in A}}
=\PRr[(x,y)\sim D]{(x,y)\in A}{{\cal X}_i}.}

For every $r\in\01^*$ let $B_r\deq T_r^{-1}(A)$.
It holds that
\m{\PR[{(x',y')\sim{D_m'}|_{{\cal X}_i}}]{T_r(x',y')\in A}
={D_m'}|_{{\cal X}_i}(B_r)
=\fr{D_m'\llp B_r\cap{\cal X}_i\rrp}{D_m'\llp{\cal X}_i\rrp},}
therefore
\m{\E[{\cR[]}\sim\U]{D_m'\llp B_r\cap{\cal X}_i\rrp}
=\fr{D_m'\llp{\cal X}_i\rrp}{D\llp{\cal X}_i\rrp}\dt D\llp A\cap{\cal X}_i\rrp.}

It is clear that $T_r$ is rectangle-invariant, so \pl[B_r] are rectangles and we can apply \lemref{razrazlem}.
\mal{-2^{-\asOm{\sqrt n}}=-2^{-\asOm m}
&\le\E[{\cR[]}\sim\U]{D_m'(B\cap\X)-\fr{D_m'(B\cap\Xo)}{135}}\\
&=\E[{\cR[]}\sim\U]{D_m'(B\cap\X)}-\fr1{135}\dt\E[{\cR[]}\sim\U]{D_m'(B\cap\Xo)}\\
&=\fr{D_m'(\X)}{D(\X)}\dt D(A\cap\X)
-\fr{D_m'(\Xo)}{135\tm D(\Xo)}\dt D(A\cap\Xo).}
Together with the facts that $D_m'(\Xo)=\fr34$ and $D_m'(\X)=\fr14$, it implies that
\mal{D(A\cap\X)
&\ge\fr{D(\X)}{135\tm D(\Xo)}\tm\fr{D_m'(\Xo)}{D_m'(\X)}\tm D(A\cap\Xo)
-\fr{D(\X)}{D_m'(\X)}\tm 2^{-\asOm{\sqrt n}}\\
&\ge\fr{D(\X)}{45}\tm D(A\cap\Xo)-2^{-\asOm{\sqrt n}}.}
Note that
\mal{
&D(\Xo)\ge\llp\fr{n-k_1(n)-k_2(n)}n\rrp^{k_2(n)}
\ge\llp1-\fr{2\alpha_2}{\sqrt n}\rrp^{\alpha_2\sqrt n}
\ge\llp\fr12\rrp^{4{\alpha_2}^2}=\llp\fr1{16}\rrp^{{\alpha_2}^2},\\ 
&D(\X)\ge k_2(n)\tm\fr{k_1(n)}n\tm D(\Xo)\ge\fr{{\alpha_1}^2}{16^{{\alpha_2}^2}}}
(the second inequality can be established analogously to the proof of \clmref{c_X}).
The result follows.}

\sect[easy_qua]{Efficient protocol for \Pin\ in \QI}
We give a 1-way quantum protocol $S$ that receives input to \Pin, communicates \asO{\log n} qubits and either produces a correct answer or refuses to answer.
For $n$ large enough the former occurs with probability at least $\fr13$.
Therefore, for any given $\eps$ one can run $t\in\asO{\log\llp\fr1{\eps}\rrp}$ instances of $S$ in parallel, thus obtaining a \f0-error protocol for \Pin\ with answering probability at least $1-\eps$.
The communication cost of the new protocol remains in \asO{\log n} as long as $\eps$ is a constant.

Let us see how $S$ works.
\enum{
\item Alice sends to Bob the state $\ket{\alpha}\deq\fr1{\sqrt n}\sum_{j\in x}\ket j$.
\item Bob measures $\ket{\alpha}$ with the $\fr{n}4+1$ projectors $E_i\deq\sum_{j\in y_i}\kbra{j}$ and $E_0\deq\sum_{j\nin\cup y_i}\kbra{j}$, let $i_0$ be the index of the outcome of the measurement and $\ket{\alpha_{i_0}}$ be the projected state.
Bob applies the Hadamard transform over $\GF2^{2\log n+1}$ to $\ket{\alpha_{i_0}}$ and measures the result in the computational basis.
Denote by $a_{i_0}$ be the outcome of the measurement.
\item If $a_{i_0}=\bar0$ or $i_0=0$ then Bob refuses to answer, otherwise he outputs $(i_0,a_{i_0})$.
}
Obviously, the protocol transmits \asO{\log n} qubits.

After the first measurement, if $i_0=0$ then Bob refuses to answer, otherwise the register remains in the state $\ket{\alpha_{i_0}}=\fr1{\sqrt2}\sum_{j\in x\cap y_{i_0}}\ket j$.
Denote by $p_i$ the probability that $i_0=i$.
Then for $i>0$,

\m{p_i=\tr\llp\kbra{\alpha}\tm E_i\rrp=
\fr1n\tm\tr\llp\llp\sum_{\mac{j,k\in x\\j\neq k}}\kbra[j]{k}+
\sum_{j\in x}\kbra{j}\rrp
\tm\sum_{j\in y_i}\kbra{j}\rrp=\fr{\sz{x\cap y_i}}n=\fr2n,}

and consequently, $p_0=1-\sum_{i>0}p_i=1/2$.

Assume that $i_0\neq0$.
Bob applies the Hadamard transform to the state $\ket{\alpha_{i_0}}=\fr{\ket{b_1}+\ket{b_2}}{\sqrt2}$ where $x\cap y_{i_0}=\set{b_1,b_2}$, denote the outcome by $\ket{\alpha_{i_0}'}$.
Then

\m{\ket{\alpha_{i_0}'}=
\fr1{2n}\tm\sum_{j\in [2n^2]}\llp(-1)^{\lra{j,b_1}}+(-1)^{\lra{j,b_2}}\rrp\ket j=
\fr1n\tm\sum_{\lra{j,b_1+b_2}=0}\pm\ket j,}

and therefore Bob obtains a uniformly random element of
\m{{\set[\lra{j,b_1+b_2}=0]{j\in [2n^2]}},}
as the outcome of his second measurement.

If $a_{i_0}=\bar0$ then Bob refuses to answer, otherwise he returns a pair $(i_0,a_{i_0})$ that satisfies the requirement.
The latter occurs with probability $1-\aso1$, conditioned on $i_0\neq0$.
So, the protocol is successful with probability $\fr12-\aso1>\fr13$, for sufficiently large $n$.

\sect[hard_cla]{Solving \Pin\ is expensive in \R}
We will establish a lower bound of $\fr{n^{\fr[]14}}{\sqrt{\log n}}$ for the 2-way classical communication complexity of \Pin.
We will always assume this model of communication, unless stated otherwise.

As outlined in \sref{s_appr}, we will first prove that solving \Pin\ implies solving \PS, then that solving \Piip\ is as simple as solving \PS, and finally that solving \Piip\ is expensive.

\ssect[hard_in2ii]{Solving \Pin\ implies solving \PS}
\lemapp{l_in2ii}
{Assume that there exists a (possibly, randomized) protocol $S$ of cost $k$ that solves \Pin\ with error bounded by $\eps$.
Then there exists $\Sigma$, such that \PS\ can be solved \wrt \Uii[2]\ with probability $2/n$ with error bounded by $2\eps$ by a deterministic protocol of cost \f{k}.}
{}
{The proof can be found in the \whereproofs.}
{\appinTOin}
{ %
Let $(x,y)$ be an instance of \PS, satisfying $|x\cap y|=2$ (recall that $x$ and $y$ are subsets of $[n^2]$, $|x|=n/2$ and $|y|=n$).
Consider the following protocol $S'$.\itemi{
\item Let $x'=\set{n^2+1\dc n^2+\fr{n}2}\cup x$.
For $1\le j\le\fr{n}4-1$, let $y_j'=\set[1\le k\le n]{n^2+j+\fr{kn}4}$ and ${\bar y}=(y,y_1'\dc y_{\fr{n}4-1}')$.
\item Using public randomness, choose random permutations:\ $\sigma_1$ over $[2n^2]$ and $\sigma_2$ over $[\fr{n}4]$.
\item Run the protocol $S$ over $\sigma_1(x', ({\bar y}_{\sigma_2(1)}\dc {\bar y}_{\sigma_2(n/4)}))$; let $(i,z)$ be the response by $S$.
\item If $\sigma_2(1)=i$ then output $(\sigma_1, z)$, otherwise refuse to answer.
}
This protocol maps the given pair $(x,y)$ to a uniformly random instance of \Pin\ (the deterministically constructed $\llp x',\bar y\rrp$ forms a correct input for \Pin, and the action of permutations upon instances of \Pin\ is transitive). 
Moreover, the original problem is mapped to a uniformly random coordinate of the instance of \Pin\ that is fed into $S$.

Let $(\cX,\cY)=(x,y)$ and $(\cX,\cY)\sim\Uii[2]$.
Denote by $\Ec$ the event that $S'$ returns an answer, by $\Ec_0$ the event that $S'$ outputs a pair $(\sigma, z)$ such that $z\neq \bar0$ and $\lra{z,a+b}=0$, where $\sigma(x\cap y)=\set{a,b}$, and by $\Ec_1$ the event $\Ec\smin \Ec_0$.
By the symmetry argument, the following holds:
If $S$ returns a correct answer then $\Ec_0$ occurs with probability $4/n$; if $S$ makes a mistake then $\Ec_1$ occurs with probability $4/n$.
In particular, $\PR{\Ec}=4/n$ and $\PR{\Ec_1}\le\eps\tm\PR{\Ec}$.

Let us derandomize $S'$.
Suppose that $S'$ uses $s$ random bits and let $\cR$ be the corresponding random variable.
Let $R_0$ be the set of $r\in\01^s$, such that $\PRr{\Ec_1}{\Ec,\cR=r}\ge2\eps$.
From the properties of $S$ it follows that
\m{\eps\tm\PR{\Ec}\ge\PR{\Ec_1}=\PR{\Ec}\tm\PRr{\Ec_1}{\Ec}\ge\PR{\cR\in R_0}\tm\PRr{\Ec}{\cR\in R_0}\tm2\eps,}
which leads to
\m{\fr12\PR{\Ec}\le\PR{\Ec}-\PR{\cR\in R_0}\tm\PRr{\Ec}{\cR\in R_0}=\PR{\cR\nin R_0}\tm\PRr{\Ec}{\cR\nin R_0}.}
Therefore, there exists some $r_0\nin R_0$, such that $\PRr{\Ec}{\cR=r_0}\ge\fr[]{\PR{\Ec}}2=\fr[]2n$ and $\PRr{\Ec_1}{\Ec,\cR=r_0}<2\eps$.

Define a deterministic protocol $S''$, which is similar to $S'$ but uses $r_0$ instead of the random string and outputs only $z$.
Observe that fixing $\cR=r_0$, in particular, fixes the mapping~$\sigma_1\deq\sigma_1'$.
Let $\Sigma$ consist of \pl[\sigma_1'], obtained as a result of the described derandomization, subsequently applied to every permitted input length.
We claim that $S''$ solves \PS\ \wrt \Uii[2]\ with probability at least $2/n$ with error bounded by $2\eps$ -- this follows from the aforementioned properties of $S'$ and the definition of $\Sigma$.
The complexity of $S''$ is $k$, as pre- and post-processing are performed locally.}

\ssect[hard_ii2iip]{Solving \PS\ implies solving \Piip}
We will show the following.
\theodup{t_ii2iip}{\theoiiTOiip}{Assume that there exists a deterministic protocol of cost $k\in\aso{n}\cap\asom1$ that solves \PS\ for some $\Sigma$ \wrt \Uii[2]\ with probability $\gamma\in\asom{2^{-k}}$ and error bounded by $10^{-22}$.
Then \Piip\ can be solved \wrt \Uii[2]\ with probability $\fr{\gamma}{k^2\tm\log^2(\fr[]n{\gamma})}$ with error \f0 by a public coin protocol of cost \asO{k+\log^2(\fr[]n{\gamma})}.}

The proof will be done in several stages.

\newcommand{\lemiiTOiipxOneExpl}{The meaning of the lemma is that a rectangle accepting input pairs from \X, but mostly those that intersect \e{not over $I_0$}, must reject, with high probability, pairs from \Xo.}

\lemapp{l_ii2iip_x1}
{Let $n$ be sufficiently large and $A$ be an input rectangle for \PS, such that $\Uii[1](A)\in2^{-\aso{n}}\cap\aso1$.
Assume that for some constant $0<\eps<1$ and $I_0\sbseq [n^2]$, $|I_0|\ge\fr{n^2}2$, it holds that
\m{\sum_{i\in I_0}\UA[1]\llp\X(i)\rrp\le\fr{8\eps}{10^7}.}
Then $\UA[0,1](\Xo)<\eps$.}
{\lemiiTOiipxOneExpl}
{\lemiiTOiipxOneExpl
The proof can be found in the \whereproofs.}
{\appiiTOiipI}
{In this proof we will casually view input pairs $(x,y)$ as \f4-tuples $(x_1,x_2,y_1,y_2)$, where $x|_{\ov{I_0}}=x_1$, $x|_{I_0}=x_2$, $y|_{\ov{I_0}}=y_1$, $y|_{I_0}=y_2$.

Let $\eps_0\deq\UA[0,1](\Xo)\in\asOm1$, in terms of this value we will derive a lower bound on the probability that a uniformly chosen \X-instance from $A$ intersects over $I_0$.

Let $(\cX,\cY)=(\cX_1,\cX_2,\cY_1,\cY_2)\sim\UA[0,1]$.
Let $x_1$ and $y_1$ be the values taken by $\cX_1$ and $\cY_1$, we define the following events characterizing these values (note that the events do not depend on the values of $\cX_2$ and $\cY_2$):\itemi{
\item $\Ec_1$ denotes the event that $\sz{x_1}\le\fr n3$ and $\sz{y_1}\le\fr{2n}3$.
\item $\Ec_2$ denotes the event that $\PRr[{\UA[0,1]}]{\Xo}{\cX_1=x_1,\cY_1=y_1}\ge\fr{\eps_0}2$.
Observe that $\Ec_2$ implies that $(x_1,y_1)\in\supp\llp\UA[0]\|_{\ov{I_0}\times\ov{I_0}}\rrp$.
\item $\Ec_3$ denotes the event that either $\lnot \Ec_2$ or $\Ec_2$ and
\m{\hh{\UA[0]}{\cX_1=x_1,\cY_1=y_1}\ge\h{\Uii[0]\|_{I_0\times I_0}}
-\llp\fr8{\eps_0}+1\rrp\tm\log\llp\fr1{\Uii[0](A)}\rrp.}
\item $\Ec_4$ denotes the event that either $\lnot \Ec_2$ or $\Ec_2$ and
\m{\hh{\Uii[0]}{\cX_1=x_1,\cY_1=y_1}\le\h{\Uii[0]\|_{I_0\times I_0}}
+\log\llp\fr8{\eps_0\tm\Uii[0,1](A)}\rrp.}
}

Our first step will be to show that all four events hold simultaneously with non-negligible probability.
This will let us apply \lemref{razlem} to many ``subrectangles'' of $A$ defined over $I_0\times I_0$, which, in turn, will lead to the desired lower bound.

The event $\Ec_1$ occurs with probability $1-2^{-\asOm n}$ if $(\cX_1,\cY_1)\sim\Uii[0,1]\|_{\ov{I_0}\times\ov{I_0}}$, due to the Chernoff bound (\clmref{c_Cher}).
In our case $(\cX_1,\cY_1)\sim\UA[0,1]\|_{\ov{I_0}\times\ov{I_0}}$, but on the other hand, $\Uii[0,1](A)\in2^{-\aso{n}}$, and therefore $\PR[{\UA[0,1]}]{\Ec_1}\in1-\aso1$.

We know that
\m{\eps_0=\UA[0,1](\Xo)=\E[{(x_1,y_1)\sim\UA[0,1]\|_{\ov{I_0}\times\ov{I_0}}}]
{\PRr[{\UA[0,1]}]{\Xo}{\cX_1=x_1,\cY_1=y_1}},}
which implies that $\PR[{\UA[0,1]}]{\Ec_2}\ge\fr{\eps_0}2$.

Let us see that $\Ec_3$ occurs with high probability.
Observe that by the chain rule,
\mal[m_x1_0]{\h{\UA[0]}&=
\h{\UA[0]\|_{\ov{I_0}\times\ov{I_0}}}+
\hh[{\UA[0]}]{\cX_2,\cY_2}{\cX_1,\cY_1},\\
\h{\Uii[0]}
&=\h{\Uii[0]\|_{\ov{I_0}\times\ov{I_0}}}+
\h{\Uii[0]\|_{I_0\times I_0}},}
where the last equality follows from the fact that \Uii[0]\ is a product distribution of its two marginal projections, as appear on the right-hand side.
Moreover, these projections are uniform over their supports, and therefore
\m[m_x1_1]{\h{\Uii[0]\|_{\ov{I_0}\times\ov{I_0}}}\ge
\h{\UA[0]\|_{\ov{I_0}\times\ov{I_0}}}}
and for any $(x_1,y_1)$ in the support of $\UA[0]\|_{\ov{I_0}\times\ov{I_0}}$,
\m[m_x1_2]{\hh[{\UA[0]}]{\cX_2,\cY_2}{\cX_1=x_1,\cY_1=y_1}\le
\h{\Uii[0]\|_{I_0\times I_0}}.}
In particular, \bref{m_x1_1} and \bref{m_x1_0} imply that
\m[m_x1_3]{\h{\Uii[0]}-\h{\UA[0]}\ge
\h{\Uii[0]\|_{I_0\times I_0}}-\hh[{\UA[0]}]{\cX_2,\cY_2}{\cX_1,\cY_1}.}

Observe that both \Uii[0]\ and \UA[0]\ are uniform over their supports; moreover, the latter support is a subset of the former.
This leads to
\m{\h{\Uii[0]}-\h{\UA[0]}=
\log\llp\fr{\sz{\supp\llp\Uii[0]\rrp}}{\sz{\supp\llp\UA[0]\rrp}}\rrp
=\log\llp\fr1{\Uii[0](A)}\rrp,}
that, together with \bref{m_x1_3}, gives us
\m{\hh[{\UA[0]}]{\cX_2,\cY_2}{\cX_1,\cY_1}\ge
\h{\Uii[0]\|_{I_0\times I_0}}-\log\llp\fr1{\Uii[0](A)}\rrp.}
Together with \bref{m_x1_2} this implies, by the Markov inequality, that $\PR[{\UA[0,1]}]{\Ec_3}\ge1-\fr{\eps_0}8$.

Let us denote by $G$ the set of pairs $(x_1,y_1)$ that falsify the condition of $\Ec_4$.
Then, starting from \bref{m_x1_0}, we get
\mal{&\h{\Uii[0]\|_{\ov{I_0}\times\ov{I_0}}}+\h{\Uii[0]\|_{I_0\times I_0}}
=\h{\Uii[0]}\ge\hh{\Uii[0]}{(\cX_1,\cY_1)\in G}\\
&\tbb=\h{\left.\llp\Uii[0]\|_{\ov{I_0}\times\ov{I_0}}\rrp\right|_G}
+\hh[{\left.\Uii[0]\right|(\cX[]_1,\cY[]_1)\in G}]{\cX_2,\cY_2}{\cX_1,\cY_1}\\
&\tbb\ge\h{\left.\llp\Uii[0]\|_{\ov{I_0}\times\ov{I_0}}\rrp\right|_G}
+\h{\Uii[0]\|_{I_0\times I_0}}+\log\llp\fr8{\eps_0\tm\Uii[0,1](A)}\rrp,}
where the last inequality is implied by the definition of $G$.
Therefore,
\m{\h{\left.\llp\Uii[0]\|_{\ov{I_0}\times\ov{I_0}}\rrp\right|_G}\le
\h{\Uii[0]\|_{\ov{I_0}\times\ov{I_0}}}-
\log\llp\fr8{\eps_0\tm\Uii[0,1](A)}\rrp.}

Both arguments of $\h{\dt}$ in the last inequality are uniform distributions over their supports, one being a subset of the other, which gives us
\mal{\log\llp\fr1{\Uii[0]\|_{\ov{I_0}\times\ov{I_0}}(G)}\rrp&=
\h{\Uii[0]\|_{\ov{I_0}\times\ov{I_0}}}-
\h{\left.\llp\Uii[0]\|_{\ov{I_0}\times\ov{I_0}}\rrp\right|_G}\\
&\ge\log\llp\fr8{\eps_0\tm\Uii[0,1](A)}\rrp.}
This leads to $\Uii[0]\|_{\ov{I_0}\times\ov{I_0}}(G)\le\fr{\eps_0}8\tm\Uii[0,1](A)$.
Note that $G$ by definition implies $\Ec_2$, thus consists exclusively of disjoint pairs, and therefore $\Uii[0,1]\|_{\ov{I_0}\times\ov{I_0}}(G)\le\Uii[0]\|_{\ov{I_0}\times\ov{I_0}}(G)\le\fr{\eps_0}8\tm\Uii[0,1](A)$.
Therefore,
\m{\PR[{\UA[0,1]}]{\Ec_4}
=1-\UA[0,1]\|_{\ov{I_0}\times\ov{I_0}}(G)\ge1-\fr{\eps_0}8.}

For $n$ sufficiently large, the events $\Ec_1$, $\Ec_3$ and $\Ec_4$ simultaneously hold with probability at least $1-\fr{\eps_0}4-\aso1>1-\fr{\eps_0}3$, and $\Ec_2$ holds with probability at least $\fr{\eps_0}2$.
The event $\Ec\deq \Ec_1\cap \Ec_2\cap \Ec_3\cap \Ec_4$ therefore holds with probability at least $\fr{\eps_0}6$ when $(\cX_1,\cY_1)\sim\UA[0,1]\|_{\ov{I_0}\times\ov{I_0}}$.

If $\Ec$ holds \wrt $\cX_1=x_1'$ and $\cY_1=y_1'$ then we can apply \lemref{razlem} to the rectangle $A_{x_1',y_1'}\deq\set[(x_1',x_2,y_1',y_2)\in A]{(x_2,y_2)}$, as follows.
Let us view $A_{x_1',y_1'}$ as an input rectangle for $\D_{\sz{I_0}}$.
Denote input to $\D_{\sz{I_0}}$ by $(\cX_2,\cY_2)$.
Define $D$ to be the distribution obtained by independently choosing $\cX_2$ and $\cY_2$ as subsets of $I_0$ of sizes $\fr n2-\sz{x_1'}$ and $n-\sz{y_1'}$, respectively.
As follows from $\Ec_1$, $\fr n6\le\sz{\cX_2}\le\fr n2$ and $\fr n3\le\sz{\cY_2}\le n$.

Observe that the mapping $M:\:I_0\times I_0\to [n^2]\times[n^2]$ defined as
\m{M(x_2, y_2)\deq(x_1'\cup x_2,\,y_1'\cup y_2)}
transforms $D$ into \Uii, conditioned upon $\cX_1=x_1',\cY_1=y_1'$.
The fact that $x_1'\cap y_1'=\emptyset$ (as implied by $\Ec_2$) means that the $M$ transforms $D$ into $\Uii|_{{\cX[]_1}=x_1',{\cY[]_1}=y_1'}$ also when the both distributions are conditioned upon some \Xj, for any valid $j$.
Note also that $M$ maps $A$ to $A_{x_1',y_1'}$.
In what follows we will be implicitly assuming equivalence between the arguments and the corresponding values of $M$, whenever necessary.

\lemref{razlem} can be applied to $A_{x_1',y_1'}$ \wrt the distribution $D$ by choosing $\alpha_1=\fr{n}{6\sqrt{\sz{I_0}}}$ and $\alpha_2=\fr{n}{\sqrt{\sz{I_0}}}$.
The conclusion is that for
$\delta=\fr{{\alpha_1}^2}{45\tm16^{{\alpha_2}^2}}\ge\fr1{207360}$,
\m[m_x1_5]{D(A_{x_1',y_1'}\cap\X)\ge\delta\tm
D(A_{x_1',y_1'}\cap\Xo)-2^{-\asOm{\sqrt{\sz{I_0}}}}
\ge\fr{D(A_{x_1',y_1'}\cap\Xo)}{207360}-2^{-\asOm n}.}

Let $D_A\deq D|_{A_{x_1',y_1'}}$ and $D_0\deq D|_{\Xo}$.
Events $\Ec_3$ and $\Ec_4$ together mean that for $n$ sufficiently large (recall that $\Uii[0](A)\in\aso1$),
\m{\hh{\UA[0]}{\cX_1=x_1',\cY_1=y_1'}\ge\hh{\Uii[0]}{\cX_1=x_1',\cY_1=y_1'}-\Delta,}
where $\Delta\in\asO{\log\llp\fr1{\Uii[0](A)}\rrp}$, as follows from $\Uii[0,1](A)\ge\Uii[0,1](\Xo)\tm\Uii[0](A)\in\asOm{\Uii[0](A)}$.
That can be restated as
\f{\h{D_A|_{\Xo}}\ge\h{D_0}-\Delta,}
and again, since the both arguments of $\h{\dt}$ are uniform distributions, one's support being a subset of the other's, this leads to
\m[m_x1_4]{D_0(A_{x_1',y_1'})\ge2^{-\Delta}=\llp\Uii[0](A)\rrp^{\asO1}.}

We know that $\Uii[1](A)\in2^{-\aso{n}}$.
Therefore, 
\m{\Uii[0,1](\Xo\cup A)=\Uii[0,1](A)\tm\UA[0,1](\Xo)=\Uii[0,1](\Xo)\tm\Uii[0](A)} together with $\Uii[0,1](\Xo)\in\asOm1$ imply
\m[m_x1_6]{\Uii[0](A)\in\asOm{\Uii[0,1](A)}\sbseq\asOm{\Uii[1](A)}\sbseq2^{-\aso{n}},}
and therefore $D_0(A_{x_1',y_1'})\in2^{-\aso{n}}$.
By the definition of $D$ it is easy to see that $D(\Xo)\in\asOm1$, and therefore
\m{D(A_{x_1',y_1'}\cap\Xo)=D(\Xo)\tm D_0(A_{x_1',y_1'})\in2^{-\aso{n}}.}
This means that for sufficiently large $n$, \bref{m_x1_5} leads to
\f{D(A_{x_1',y_1'}\cap\X)>D(A_{x_1',y_1'}\cap\Xo)/207361,} implying $\PRr[D_A]{\X}{\Xo\cup\X}>\fr1{207361}$.

We know that $\Ec$ occurs with probability at least $\fr{\eps_0}6$, and thus
\m{\sum_{i\in I_0}\UA[1]\llp\X(i)\rrp\ge
\sum_{i\in I_0}\UA[0,1]\llp\X(i)\rrp\ge
\PR[{\UA[0,1]|_{\ov{I_0}\times\ov{I_0}}}]{\Ec}
\tm\mathbf{Pr^*}[\X|\Xo\cup\X]>\fr{8\eps_0}{10^7},}
where $\mathbf{Pr^*}[\X|\Xo\cup\X]$ denotes the maximum possible value of $\PRr{\X}{\Xo\cup\X}$, taken over $D_A$ that is defined as above, over some pair $x_1',y_1'$ for which $\Ec$ holds.
The result follows.}

We will need the following extension of \lemref{l_ii2iip_x1} to the case of rectangles, selectively accepting instances of $\Xii$.

\newcommand{\lemiiTOiipxTwoExpl}{An alternative way to look at $\set{I_0^{(i)}}$ would be to say that $I_0^{(i)}$ is the set of neighbors of $i$ in an undirected graph without self-loops over $n^2$ vertices, of degree at least $\fr[]{n^2}2$ each.
Call that graph $\Gamma$, then the requirement of the lemma is that a randomly chosen pair of vertices $\set{i,j}=\cX\cap\cY$ when $(\cX,\cY)\sim\UA[2]$ is unlikely to be connected in $\Gamma$.
If the condition is met (\ie such $\Gamma$ can be found), then $A$ cannot be large.}

\lemapp{l_ii2iip_x2}
{Let $n$ be sufficiently large and $A$ be an input rectangle for \PS, such that $\Uii[2](A)\in2^{-\aso{n}}$.
Let $\set{I_0^{(i)}}_{i\in [n^2]}$ be a family of subsets of $[n^2]$, such that for every $i,j\in [n^2]$ it holds that $i\nin I_0^{(i)}$, $|I_0^{(i)}|\ge\fr{n^2}2$, and $i\in I_0^{(j)}$ if and only if $j\in I_0^{(i)}$.
If $A$ satisfies that
\m{\fr12\sum_{\mac{i\in [n^2]\\j\in I_0^{(i)}}}\UA[2]\llp\Xii(i,j)\rrp\le\fr{\eps^2}{10^{20}},}
then $\UA[0,1,2](\Xo\cup\X)<\eps$.}
{\lemiiTOiipxTwoExpl}
{\lemiiTOiipxTwoExpl\
The proof can be found in the \whereproofs.}
{\appiiTOiipII}
{We will show that $\UA[0,1,2](\X)<\fr{\eps}{2883}$ and $\UA[0,1,2](\Xo)<\fr{2881}{2883}\eps$.

Define $A_i\deq\set[i\in x\cap y]{(x,y)\in A}$ for each $i\in [n^2]$.
Let $D$ be the probability distribution over $[n^2]$ defined by $D(i)=\fr12\UA[2](A_i)$.
Choosing $(\cX,\cY)\sim\UA[2]$ can be viewed as first choosing $i\sim D$, followed by $(\cX,\cY)\sim\U[A_i]^{(2)}$,
and our assumptions guarantee that 
\m{\E[i\sim D]{\sum_{j\in I_0^{(i)}}\U[A_i]^{(2)}(\Xii(i,j))}\le\fr{\eps^2}{10^{20}}.}

Let 
\mal{
&I_1\deq\set[{\Uii[2](A_i)<\fr{\eps}{10^7\tm n^2}\tm\Uii[2](A)}]{i\in [n^2]},\\
&I_2\deq
\set[{\sum_{j\in I_0^{(i)}}\U[A_i]^{(2)}(\Xii(i,j))>
\fr{\eps}{10^{12}}}]{i\in [n^2]}.}
Then
\m{\sum_{i\in I_1}\Uii[2](A_i)<\fr{\eps}{10^7}\tm\Uii[2](A)\,\Rightarrow\,
\sum_{i\in I_1}\UA[2](A_i)<\fr{\eps}{10^7},}
and by Markov inequality,
\m{D(I_2)<\fr{\eps}{10^8}\,\Rightarrow\,
\sum_{i\in I_2}\UA[2](A_i)<\fr{2\eps}{10^8}.}
That is,
\m[m_x2_0]{\sum_{i\in I_1\cup I_2}\UA[2](A_i)
<\fr{1.2\,\eps}{10^7}.}

For any $i_0\in [n^2]$, we view $A_{i_0}$ as an input rectangle for \PS, defined over $[n^2]\mset{i_0}$.\fn
{Strictly speaking, this violates our requirement that $n$ is a power of $2$ and slightly affects the Hamming weights of $x$ and $y$ as functions of $n$.
However, the former is irrelevant in this context and the influence of the latter is negligible for sufficiently large $n$, so we allow this abuse to keep the notation simple.
Note also that we keep counting the size of $x\cap y$ when using the \Xj-notation according to the original definition of the communication task, \ie \wrt \PS\ defined over $[n^2]$.}
Assume $i_0\in [n^2]\smin I_1\smin I_2$, then it holds that $\Uii[2](A_{j_0})\ge\fr{\eps}{10^7\tm n^2}\tm\Uii[2](A)\in2^{-\aso{n}}$.
The properties of $I_0^{(i_0)}$ and the fact that $i_0\nin I_2$ allow us to apply \lemref{l_ii2iip_x1}, concluding that \m[m_x2_1]{\U[A_{i_0}]^{(1,2)}(\X)<\fr{\eps}{8\tm10^5}.}

For every $i_0\in I_1\cup I_2$ we, on the other hand, apply \lemref{razlem} to $A_{i_0}$ (replacing \Xo\ by \X\ and \X\ by \Xii, due to the guaranteed $i_0\in(x,y)$ for every $(x,y)\in A_{i_0}$).
Then for $\delta=\fr1{2880}$ (we use $\alpha_1=1/2$ and $\alpha_2=1$, in accordance with the definition of \PS),
\m[m_x2_3]{\sum_{i\in I_1\cup I_2}\Uii(A_i\cap\X)\le
\fr1{\delta}\tm\sum_{i\in I_1\cup I_2}\Uii(A_i\cap\Xii)+n^2\tm2^{-\asOm{n}}.}
Clearly,
\m[m_x2_4]{\Uii\llp(\X\cup\Xii)\cap A\rrp\ge\Uii(\Xii\cap A)=
\Uii(\Xii)\tm\Uii[2](A)\in2^{-\aso{n}},}
and dividing \bref{m_x2_3} by $\Uii\llp(\X\cup\Xii)\cap A\rrp$ gives
\mal[m_x2_2]{&\sum_{i\in I_1\cup I_2}\UA[1,2](A_i\cap\X)\le
\fr1{\delta}\tm\sum_{i\in I_1\cup I_2}\UA[1,2](A_i\cap\Xii)+
2^{-\asOm{n}}\\
&\tbb\le\fr1{\delta}\tm\sum_{i\in I_1\cup I_2}\UA[2](A_i)+2^{-\asOm{n}}
\le\fr{1.2\,\eps}{\delta\tm10^7}+2^{-\asOm{n}},}
as follows from \bref{m_x2_0}.

We conclude that for sufficiently large $n$,
\mal{\UA[0,1,2](\X)
&\le\UA[1,2](\X)
=\sum_{i\in [n^2]}\UA[1,2](A_i\cap\X)\\
&=\sum_{i\in I_1\cup I_2}\UA[1,2](A_i\cap\X)+
\sum_{i\nin I_1\cup I_2}\UA[1,2](A_i)\tm\U[A_{i_0}]^{(1,2)}(\X)\\
&<\fr{1.2\,\eps}{\delta\tm10^7}+2^{-\asOm{n}}+\fr{\eps}{8\tm10^5}
<\fr{\eps}{2883},}
as follows from \bref{m_x2_1} and \bref{m_x2_2}.

We apply \lemref{razlem} one more time.
For the same value of $\delta$ it holds that
\m{\Uii(A\cap\Xo)\le\fr1{\delta}\tm\Uii(A\cap\X)+2^{-\asOm{n}}.}
Like in \bref{m_x2_4},
\m{\Uii\llp(\Xo\cup\X\cup\Xii)\cap A\rrp\ge\Uii(\Xii\cap A)\in2^{-\aso{n}},}
and therefore for sufficiently large $n$,
\m{\UA[0,1,2](\Xo)\le\fr1{\delta}\tm\UA[0,1,2](\X)+2^{-\asOm{n}}
<\fr{2880}{2883}\,\eps+2^{-\asOm{n}}<\fr{2881}{2883}\,\eps,}
as required.}

In order to state our next lemma we need the following definition.
\defi{We call a rectangle $A$ \e{$\delta$-labeled} if
\m{\PR[{\cY[]}\sim\UAb]
{\exists\, a,b\in \cY,\, a\neq b:\:\PR[{\cX[]}\sim\UAa]{\set{a,b}\sbs \cX}\ge\delta}>\fr13.}
}

The following lemma (which is our final step towards \theoref{t_ii2iip}) claims, informally, that if a big rectangle $A$ accepts instances of $\Xii$ while rejecting those of $\Xo$ and $\X$, then there can be only limited uncertainty regarding the content of $\cX\cap\cY$, for a randomly chosen pair $(\cX,\cY)\in A$.
\lemapp{l_ii2iip_3}
{Let $n$ be sufficiently large and $A$ be an input rectangle for \PS, such that $\PR[\UA]{\sz{\cX\cap\cY}<2}\le\fr16$ and $A$ is not $\delta$-labeled for some $\delta>0$.
Then $\Uii(A)\in2^{-\asOm{\fr1{\sqrt{\delta}}}}$.}
{}
{The proof can be found in the \whereproofs.}
{\appiiTOiipIII}
{Let $(\cX,\cY)$ be the input variables.
Let $B$ be the set of $y\sbs [n^2]$, such that
\m[m_cond1]{\PR[{\cX[]}\sim\UAa]{\sz{\cX\cap y}<2}\le\fr13}
and 
\m[m_cond2]{\forall\, a,b\in y,\, a\neq b:\:\PR[{\cX[]}\sim\UAa]{\set{a,b}\sbs\cX}<\delta.}
If we choose $y=\cY\sim\UAb$ then \bref{m_cond1} holds with probability at least $\fr12$ and \bref{m_cond2} holds with probability at least $\fr23$ ($A$ is not $\delta$-labeled); therefore, $\Uii(A')\ge\fr16\Uii(A)$ for $A'\deq A\sA\times B$.

For $a,b\in [n^2]$, $a\neq b$, let $p_a\deq\PR[{\cX[]}\sim\UAa]{a\in\cX}$ and $p_b^{(a)}\deq\PRr[{\cX[]}\sim\UAa]{b\in\cX}{a\in\cX}$.
Condition \bref{m_cond2} implies that
\m[m_cond3]{\forall a\in y:\:\llp p_a\ge\sqrt{\delta}\,\Rightarrow\,
\forall b\in y\mset a:\:p_b^{(a)}<\sqrt{\delta}\rrp.}

We will see that both \bref{m_cond1} and \bref{m_cond3} are not likely to hold simultaneously for a random $y=\cY\sim\Ub$.
Let $a_0(y)$ denote the lexicographically first value $i_0\in y$, satisfying $p_{i_0}=\Max[i\in y]{p_i}$.

First let us consider the situation when $p_{a_0({\cY[]})}<\sqrt{\delta}$, \ie
\m[m_cond4]{\forall a\in y:\:p_a<\sqrt{\delta}.}
Since \bref{m_cond1} implies $\sum_{a\in\cY[]}p_a\ge\fr43$, the probability of both \bref{m_cond1} and \bref{m_cond4} holding simultaneously \wrt $y=\cY$ is upper bounded by
\m[m_cond5]{\PR{\sum_{a\in\cY[]}p'_a\ge\fr43},}
where $p'_a\deq\twocase{p_a}{if $p_a<\sqrt{\delta}$}{0}{otherwise}$.

Let $Z_1\dc Z_n$ be the elements of $\cY$ and denote $W_i\deq p'_{Z_i}$.
We want to use Chernoff bound in order to limit from above the value of $\sum_{i=1}^nW_i$.
Though the variables $W_i$ are not independent (because all \pl[Z_i] must be different), it is possible to apply Chernoff bound using the ``worst case'' estimation of the variables' conditional mean values.
Formally, in order to obtain the lower bound we analyze a relaxation of the original experiment, where all \pl[W_i] are independent but distributed according to the worst scenario, resulting from conditioning upon the values of \set[j\ne i]{W_j}.
Note that for each $1\le i\le n$ it holds that $W_i\le\sqrt{\delta}$ and $\E{W_i}\le\fr{\sz x}{\sz{[n^2]}-\sz y}=\fr{n/2}{n^2-n}<\fr3{5n}$, even if the expectation is conditioned upon some values of \set[j\ne i]{W_j}.
Based on Chernoff bound (\clmref{c_Cher}), we conclude that
\m[m_bound_on5]{\PR[{\cY[]}\sim\Ub]{\sum_{a\in \cY[]}p'_a\ge\fr43}\in
2^{-\asOm{\fr1{\sqrt{\delta}}}}.}

Now consider the other choice left by \bref{m_cond3}, namely
\m[m_cond6]{p_{a_0}\ge\sqrt{\delta}\txt{~~and~~}
\forall b\in\cY, b\neq a_0:\:p_b^{(a_0)}<\sqrt{\delta},}
where $a_0$ stands for $a_0(\cY)$.
Let $\cY=y$; since $\PR[\UAa]{\sz{\cX\cap y}\ge2}\ge\fr23$ implies
\m{\sum_{b\in y\mset{a_0}}p_b^{(a_0)}\ge\fr23,}
the probability that \bref{m_cond1} and \bref{m_cond6} hold is upper bounded by the probability that
\m[m_cond7]{\sum_{b\in y\mset{a_0}}{p_b^{(a_0)}}'\ge\fr23,}
where ${p_b^{(a_0)}}'\deq\twocase{p_b^{(a_0)}}{if $p_b^{(a_0)}<\sqrt{\delta}$}{0}{otherwise}$.

Like in the case of \bref{m_cond5}, Chernoff bound (\clmref{c_Cher}) implies that \bref{m_cond7} holds with probability $2^{-\asOm{\fr1{\sqrt{\delta}}}}$.
Therefore,
\m{\Uii(A)\le6\dt\Uii(A')\le
6\dt\PR[{\cY[]}\sim\Ub]{\cY\in B}\in2^{-\asOm{\fr1{\sqrt{\delta}}}},}
as required.}

\newcommand{\prfTiiTOiip}{\prf[\theoref{t_ii2iip}]
{Let $S$ be a deterministic protocol of cost $k$ solving \PS\ for some $\Sigma$ \wrt \Uii[2]\ with probability $\gamma$ and error bounded by $10^{-22}$.

Let $A$ be an input rectangle; observe that \lemref{l_ii2iip_3} guarantees that if $\UA(\Xo\cup\X)\le\fr16$ and $\Uii(A)\ge2^{-\asOm k}$ then there exists a function $\delta(k)\in\asOm{\fr1{k^2}}$, such that $A$ is $\delta(k)$-labeled.
Fix such $\delta(k)$ for the rest of the proof.

Consider the rectangles defined by $S$.
We will call a rectangle $A$ \e{latent} if it is not possible to define an answer that would solve \PS\ with probability at least $1-\fr2{10^{22}}$ \wrt \UA[2].
As follows from the accuracy of $S$, $(\cX,\cY)\sim\Uii[2]$ does not belong to a latent rectangle with probability at least $\fr{\gamma}2$ (at least half of all pairs $(x,y)\in\Xii$ for which $S$ produces an answer belong to non-latent rectangles, since otherwise the error of $S$ would be greater than  $10^{-22}$).
On the other hand, with probability at least $1-\fr{\gamma}4$ it happens that $(\cX,\cY)\sim\Uii[2]$ falls into a rectangle $A$ satisfying $\Uii[2](A)\ge\fr{\gamma}{2^{k+2}}$.
Note that for any such $A$ it holds that $\Uii[2](A)\ge2^{-2k}$ and $\Uii(A)\ge\Uii(\Xii)\tm\fr{\gamma}{2^{k+2}}\ge2^{-2k}$ (if $n$ is large enough).

Call a rectangle $A$ \e{good} if it is not latent, $\Uii[2](A)\ge2^{-2k}$ and $\Uii(A)\ge2^{-2k}$.
As shown above, $(\cX,\cY)\sim\Uii[2]$ belongs to a good rectangle with probability at least $\fr{\gamma}2-\fr{\gamma}4=\fr{\gamma}4$.
Consequently, $(\cX,\cY)\sim\Uii[\ge2]$ falls into a good rectangle with probability at least $\Uii(\Xii)\tm\fr{\gamma}4\ge\fr{\gamma}{52}$.

We claim that any good $A$ is $\delta(k)$-labeled.
This follows from the fact that there exists some $z_A\in [n^2]\mset{0}$, such that
\m{1-\fr2{10^{22}}\le\PR[{\UA[2]}]{(\cX,\cY,z_A)\in\PS}
=\PR[{\UA[2]}]{\lra{z_A,\sigma_{n^2}(a)+\sigma_{n^2}(b)}=0},}
where $\cX\cap\cY=\set{a,b}$ and $\sigma_{n^2}\in\Sigma$.
If we define $I_0^{(a)}\deq\set[\lra{z_A,\sigma_{n^2}(a)+\sigma_{n^2}(b)}=1]{b\in [n^2]}$ that will, \wrt $A$, satisfy the requirement of \lemref{l_ii2iip_x2} for $\eps=\fr16$.
Therefore it holds that $\UA(\Xo\cup\X)\le\UA[0,1,2](\Xo\cup\X)<\fr16$.
As $\Uii(A)\ge2^{-2k}$, we can apply the contrapositive of \lemref{l_ii2iip_3}, as suggested in the beginning of the proof, which leads to the conclusion that $A$ is $\delta(k)$-labeled.

We are ready to construct a protocol, as promised by the statement we are proving.
The idea is to first map the input $(x,y)\in\Xii$ to $(x',y')=(\cX',\cY')\sim\Uii[\ge2]$, then to feed $(x',y')$ to the original protocol $S$, hoping that the pair will fall into a $\delta(k)$-labeled rectangle.
If that occurs, we have a good candidate for the correct answer, as guaranteed by the fact that $A$ is $\delta(k)$-labeled.
Validity of the guess is easy to verify by a 2-way protocol.

Let $D$ be the distribution over $[n]$ defined by $D(j)\deq\Uii[\ge2](\Xj)$.
Consider the following protocol $S'$.
\enum{
\item \label{pro_ch1} Alice and Bob use public randomness to choose $j_0\sim D$.
If $j_0>3\log\llp\fr{312}{\gamma\tm\delta(k)}\rrp$ then the protocol stops and returns no answer.
Otherwise Alice sends to Bob $j_0$ lexicographically first elements from $x$, denoted by $(x_1\dc x_{j_0})$.
\item Bob sends to Alice any two indices $i_1$ and $i_2$, such that $I_x\deq\set{x_i}_{i=1}^{j_0}\mset{x_{i_1}, x_{i_2}}$ and $y$ are disjoint, followed by $j_0$ lexicographically first elements from $y$, denoted by $(y_1\dc y_{j_0})$.
\item Let $i_3$ and $i_4$ be any two indices, such that $I_y\deq\set{y_i}_{i=1}^{j_0}\mset{y_{i_3}, y_{i_4}}$ and $x$ are disjoint, denote $\tilde x\deq\llp x\cup I_y\rrp\smin I_x$.
\item Alice and Bob use public randomness to choose a random permutation $\rho$ over the elements of $[n^2]$.
\item \label{pro_ch2} \label{pro_Sran} Alice and Bob run the protocol $S$ on the input $(\rho(\tilde x),\rho(y))$.
Let $A$ be the rectangle defined by $S$, where $(\rho(\tilde x),\rho(y))$ belongs.
If there exists no pair $a,b\in y,\, a\neq b$, such that $\PR{\set{a,b}\sbs\cX}\ge\delta(k)$ when $\cX\sim\UAa$, then the protocol stops and returns no answer; otherwise let $(a',b')$ be any such pair.
\item \label{pro_ch3} If $\set{\rho^{-1}(a'), \rho^{-1}(b')}\sbseq x\cap y$ then the protocol outputs these two elements.
Otherwise the protocol returns no answer.
}

It is clear that the protocol is \f0-error and its communication cost is $\asO{k+j_0\tm\log n}\sbseq\asO{k+\log^2(\fr[]n{\gamma})}$.
Let us calculate the probability that an answer is produced.

Consider an ``idealized'' protocol $S''$, similar to $S'$ but having no halting condition in stage $1$ (\ie $S''$ continues to run regardless of the value of $j_0$).
Define the following events characterizing the behavior of $S''$:\itemi{
\item[-] $\Ec_1$ is the event that $(\rho(\tilde x),\rho(y))$ belongs to a $\delta(k)$-labeled rectangle $A$.
\item[-] $\Ec_2$ is the event that at step \ref{pro_ch2} a pair $(a',b')$ has been chosen.
\item[-] $\Ec_3$ is the event that $\Ec_2$ occurs and $\set{a',b'}\sbs x$.
\item[-] $\Ec_4$ is the event that $\Ec_3$ occurs and $j_0\le3\log\llp\fr{312}{\gamma\tm\delta(k)}\rrp$.
\item[-] $\Ec_5$ is the event that $\Ec_4$ occurs and $\set{\rho^{-1}(a'), \rho^{-1}(b')}=x\cap y$.
}
Clearly, the probability that $S'$ is successful is equal to the probability that $\Ec_5$ occurs.

Note that since $\rho$ is a uniformly random permutation and $j_0\sim D$, it holds that $(\rho(\tilde x),\rho(y))\sim\Uii[\ge2]$, and so $\PR{\Ec_1}\ge\fr{\gamma}{52}$.
By the definition of a $\delta(k)$-labeled rectangle, $\PRr{\Ec_2}{\Ec_1}\ge\fr13$ and $\PRr{\Ec_3}{\Ec_2}\ge\delta(k)$, so $\PR{\Ec_3}\ge\fr{\gamma\tm\delta(k)}{156}$.

Event $\Ec_4$ occurs if $\Ec_3$ occurs and $j_0\le3\log\llp\fr{312}{\gamma\tm\delta(k)}\rrp$, therefore 
\m{\PR{\Ec_4}
\ge\fr{\gamma\tm\delta(k)}{156}
-\PR[D]{j_0>3\log\llp\fr{312}{\gamma\tm\delta(k)}\rrp}
\ge\fr{\gamma\tm\delta(k)}{312},}
where the second inequality follows from \clmref{c_X}.

Finally, $\Ec_5$ occurs if $\Ec_4$ occurs and the points $\rho^{-1}(a')$ and $\rho^{-1}(b')$ belong to $x\cap y$.
Given $j_0$, our mapping of $(x,y)$ to $(\rho(\tilde x),\rho(y))$ produces a random instance drawn from $\Uii[\ge2](\Xj)$.
Moreover, the two elements of $x\cap y$ are mapped to a uniformly random pair inside $\rho(\tilde x)\cap \rho(y)$, even if we condition upon $\Ec_4$ (the pair $(\rho(\tilde x), \rho(y))$ is the input to $S$ at step \ref{pro_Sran}, and it reveals no additional information about $\rho(x\cap y)$ inside $\rho(\tilde x\cap y)$).
The conditional probability that $\set{\rho^{-1}(a'), \rho^{-1}(b')}=x\cap y$ is equal to $\fr[]1{\chs{j_0}2}\ge\fr1{j_0^2}$, and
\m{\PR{\Ec_5}=\PR{\Ec_4}\tm\PRr{\Ec_5}{\Ec_4}\ge
\fr{\gamma\tm\delta(k)}{312\tm j_0^2}\in
\asOm{\fr{\gamma}{k^2\tm\log^2(\fr[]n{\gamma})}},}
as follows from $\delta(k)\in\asOm{\fr[]1{k^2}}$.

The protocol $S'$ is \f0-error, so we can repeat it several times in order to get an answer with probability at least $\fr{\gamma}{k^2\tm\log^2(\fr[]n{\gamma})}$.}}

\newcommand{\aboutThiiTOiip}{In the proof we, essentially, argue that a protocol solving \PS\ \wrt \Uii[2]\ must give rise to ``typical'' rectangles satisfying the requirements of \lemref{l_ii2iip_x2}, which lets us apply the contrapositive of \lemref{l_ii2iip_3} and conclude that typical rectangles are $\delta$-labeled.
From the definition of $\delta$-labeled, the pair \set{a,b}\ chosen \wrt \cY\ has good chances to equal $\cX\cap\cY$, if the input belongs to that rectangle.
The last observation leads to a protocol for solving \Piip.}

Recall \theorep{t_ii2iip}{\theoiiTOiip}
Now we have all that is required to prove it.
\aboutThiiTOiip\
\prfTiiTOiip 

\ssect[hard_iip]{Solving \Piip\ is expensive}
It is not hard to see that a protocol of communication cost $k$ can solve \Piip\ \wrt \Uii[1] only with probability \asO{\fr kn}.
We prove the following generalization of this statement.
\theo[t_hard_iip]{Let $t\in\aso{\sqrt n}$, then any \f0-error public coin protocol of cost $k\in\asOm{t\log n}$ solving \Piip\ \wrt \Uii[t] can succeed with probability $\llp\asO{\fr{kt}n}\rrp^t$.}

This is a direct product theorem, because its statement can be rephrased as one about solving $t$ independent instances of \Piip\ \wrt \Uii[1].
There are known direct product results that apply to problems like \D\ and \Piip\ (\eg \cite{JKN08_Dir}, and references therein).
However, two obstacles prevent us from using them: on the one hand, those results mostly apply to the case of product input distributions; on the other hand, their statements have ``\asOm{t}'' in the exponent of the guaranteed upper bound on success probability, and that is not sufficient to us.
In \sref{s_open} we pose some related open questions.

In this paper we will only make use of the case corresponding to $t=2$, though we prove the theorem in full generality, as it might be of independent interest.

\prf[\theoref{t_hard_iip}]{Let $S$ be a \f0-error protocol of cost $k$ solving \Piip\ \wrt \Uii[t] with probability $\pt$.
For $i>t$, let $\pt[i]$ be the probability that $S$ outputs $t$ elements from $x\cap y$ when $(x,y)=(\cX,\cY)\sim\Uii[i]$.

\prp*{There exists an absolute constant $c$ such that for $t\le i\le\fr n2$ it holds that
\m{\pt[i]\le
\Max{\llp\fr kn\rrp^t,\llp1+\fr{ck}n\rrp\tm\llp1-\fr{t}{i+1}\rrp\tm\pt[i+1]}.}
}

The proposition implies the theorem, as follows.
Let $n$ be such that $t+\floor{\fr{n}{3ck}}<\fr n2$.
If for any $i\in\set{t\dc t+\floor{\fr{n}{3ck}}}$ it holds that $\pt[i]\le\llp\fr kn\rrp^t$, let $i_0$ be the smallest value like this, then
\m{\pt\le\llp1+\fr{ck}n\rrp^{i_0-t}\pt[i_0]\in\asO{\llp\fr kn\rrp^t}.}
Otherwise,

\m{\pt\le\llp1+\fr{ck}n\rrp^{\floor{\fr{n}{3ck}}}\tm
\prod_{i=t}^{t+\floor{\fr{n}{3ck}}-1}\fr{i+1-t}{i+1}\tm\pt[t+\floor{\fr{n}{3ck}}]\le
2\fr{\prod_{i=1}^{t}i}{\prod_{j=\floor{\fr{n}{3ck}}+1}^{t+\floor{\fr{n}{3ck}}}j}
\in\llp\asO{\fr{kt}n}\rrp^t.}

Let us prove the proposition.
Let $i_0\in\set{t\dc\fr{n}2}$ be such that $\pt[i_0]>\llp1-\fr{t}{i_0+1}\rrp\pt[i_0+1]$ and $\pt[i_0]>\llp\fr kn\rrp^t$, our goal is to show that $\pt[i_0]\le\llp1+\fr{ck}n\rrp\llp1-\fr{t}{i+1}\rrp\pt[i+1]$ for some fixed $c$.

Let $m\deq n^2-i_0$, define $D$ as the uniform distribution over pairs $(x',y')$ \st $x'\sbs[m]$, $\sz{x'}=n/2-i_0$, $y'\sbs[m]$ and $\sz{y'}=n-i_0$.
Assume we know that $(x',y')\in\supp(D)$ belongs to either \Xo\ or \X, and want to distinguish the two cases.
Consider the following public coin protocol $S'$, running on $(x',y')$.
\enum{
\item Let $x'_0\deq x'\cup\set{j}_{j=m+1}^{n^2}$ and $y'_0\deq y'\cup\set{j}_{j=m+1}^{n^2}$.
Alice and Bob use public randomness to choose a random permutation $\rho$ over the elements of $[n^2]$.
\item Alice and Bob run the protocol $S$ on the input $(\rho(x'_0),\rho(y'_0))$.
If $S$ does not outputs $t$ elements then $S'$ refuses to answer.
Otherwise if the $t$ produced elements belong to $\rho(\set[m<j\le n^2]{j})$ then $S'$ outputs \fb{0}, else $S'$ refuses to answer.
}

If $(x',y')\in\Xo$ then the pair $(\rho(x'_0),\rho(y'_0))$ is distributed according to \Uii[i_0] and $S'$ outputs \fb{0} with probability $\pt[i_0]$.
If $(x',y')\in\X$ then the pair $(\rho(x'_0),\rho(y'_0))$ is distributed according to \Uii[i_0+1] and $S'$ outputs \fb{0} with probability $\pt[i_0+1]\tm\fr[]{\chs{i_0}t}{\chs{i_0+1}t}=\llp1-\fr{t}{i_0+1}\rrp\tm\pt[i_0+1]$.
We know that the former probability is higher than the latter, and so if  $S'$ outputs \fb{0} that can be viewed as an argument towards $(x',y')\in\Xo$.

Note that $D(\Xo)\ge\fr13$ (by analogy to \clmref{c_X}), and we can apply \lemref{razlem} with $\alpha_1=\fr14$ and $\alpha_2=1$.
The lemma implies that for $\delta=\fr1{11520}$, some absolute constant $c_0$ and any rectangle $A$ it holds that \m[m_hard_iip_1]{D(A\cap\X)\ge\delta\tm D(A\cap\Xo)-2^{-c_0\tm n}.}

Let $l\in\NN$ and $S_l'$ be a protocol that runs $S'$ as a subroutine $l$ times (each time using independent random bit), and outputs \fb{0} if all the instantiations of $S'$ return \fb{0} (otherwise $S_l'$ refuses to answer).
Denote by $\Ec_0$ the event that $S_l'$ outputs \fb{0}.
If $(x',y')\in\Xo$ then $\Ec_0$ occurs with probability $\llp\pt[i_0]\rrp^l$, if $(x',y')\in\X$ then $\Ec_0$ occurs with probability $\llp\llp1-\fr{t}{i_0+1}\rrp\tm\pt[i_0+1]\rrp^l$.
Therefore, \wrt uniformly random bits used by $S_l'$, we expect that
\m{\PR[D]{\Xo\txt{ and }\Ec_0}\ge\fr13\tm\llp\pt[i_0]\rrp^l}
and
\m{\PR[D]{\X\txt{ and }\Ec_0}\le\llp\llp1-\fr{t}{i_0+1}\rrp\tm\pt[i_0+1]\rrp^l.}

Assume that $S_l'$ uses $s$ random bits, for any $r\in\01^s$ let $S_l'(r)$ be the deterministic protocol obtained from $S_l'$ by using $r$ instead of the random bits.
Because $S_l'(r)$ is a protocol of communication cost $kl$, it partitions the domain into rectangles $A_1^{(r)}\dc A_{2^{kl}}^{(r)}$.
Let $B$ consist of all \pl[A_i^{(r)}] for $r\in\01^s$, on which the corresponding $S_l'(r)$ outputs \fb{0}.
We denote
\m{\beta(l)\deq
\fr13\dt\llp\fr{\pt[i_0]}{\llp1-\fr{t}{i_0+1}\rrp\tm\pt[i_0+1]}\rrp^l\le
\fr{\PR[D]{\Xo\txt{ and }\Ec_0}}{\PR[D]{\X\txt{ and }\Ec_0}},}
then 

\m{\fr1{2^s}\tm\sum_{A\in B}D\llp A\cap\Xo\rrp
=\PR[D]{\Xo\txt{ and }\Ec_0}\ge\beta(l)\tm\PR[D]{\X\txt{ and }\Ec_0}
=\fr{\beta(l)}{2^s}\tm\sum_{A\in B}D\llp A\cap\X\rrp.}

Let $\mu\deq\E[A\in B]{D\llp A\cap\Xo\rrp}$ and $B'\deq\set[D\llp A\cap\Xo\rrp\ge\fr{\mu}2]{A\in B}$.

Then

\m{\sum_{A\in B'}D\llp A\cap\Xo\rrp
\ge\fr12\sum_{A\in B}D\llp A\cap\Xo\rrp
\ge\fr{\beta(l)}2\sum_{A\in B}D\llp A\cap\X\rrp
\ge\fr{\beta(l)}2\sum_{A\in B'}D\llp A\cap\X\rrp,}

and there exists $A_0\in B'$ satisfying 
\m{\fr2{\beta(l)}D\llp A_0\cap\Xo\rrp\ge D\llp A_0\cap\X\rrp.}

It holds that

\m{\mu\ge\fr1{2^{kl}}\tm\PR[D]{\Xo\txt{ and }\Ec_0}
\ge\fr{\llp\pt[i_0]\rrp^l}{3\tm2^{kl}}
>\fr{k^{tl}}{3\tm2^{kl}\tm n^{tl}}
>2^{-kl-tl\log n-2},}

and $D\llp A_0\cap\Xo\rrp\ge\fr{\mu}2>2^{-kl-tl\log n-3}$.
So, \bref{m_hard_iip_1} leads to

\mal{
&\fr2{\beta(l)}\tm D\llp A_0\cap\Xo\rrp\ge D\llp A_0\cap\X\rrp
\ge\delta\tm D(A_0\cap\Xo)-2^{-c_0\tm n};\\
&2^{-c_0\tm n}\ge\llp\delta-\fr2{\beta(l)}\rrp\tm D(A_0\cap\Xo)
\ge\llp\delta-\fr2{\beta(l)}\rrp\tm2^{-kl-tl\log n-3};\\
&\delta-\fr2{\beta(l)}\le2^{l(k+t\log n)+3-c_0\tm n}.}

Recall that $k\in\asOm{t\log n}$, so there exists an absolute constant $c_1$ that guarantees that the right-hand side of the last inequality is less than $\fr{\delta}2$, as long as $l\le\fr{c_1n}k$.
Consequently,
\m{\fr4{\delta}\ge\beta\llp\fr{c_1n}k\rrp=\fr13\dt
\llp\fr{\pt[i_0]}{\llp1-\fr{t}{i_0+1}\rrp\tm\pt[i_0+1]}\rrp^{\fr{c_1n}k},}
which implies that for some absolute constant $c$,

\m{\llp\fr{\pt[i_0]}{\llp1-\fr{t}{i_0+1}\rrp\tm\pt[i_0+1]}\rrp^{\fr nk}\le c
\,\Rightarrow\,\fr{\pt[i_0]}{\llp1-\fr{t}{i_0+1}\rrp\tm\pt[i_0+1]}\le1+\fr{ck}n,}

as required.}

\ssect[hard_in]{Lower bound on the classical 2-way communication complexity of \Pin}
\theo[t_main_low]{Solving \Pin\ in the classical 2-way setting with bounded error requires a protocol of cost \asOm{\fr{n^{\fr[]14}}{\sqrt{\log n}}}.}

\prf[\theoref{t_main_low}]{
Assume that a protocol $S$ of communication cost $k\in\aso n$ solves \Pin\ with error bounded by $\fr1{2\tm10^{22}}$.

Then \lemref{l_in2ii} implies that there exists a protocol $S'$ of communication cost \asO k\ that solves \PS\ for some $\Sigma$ \wrt \Uii[2]\ with probability $\fr2n$ and error bounded by $\fr1{10^{22}}$.

By \theoref{t_ii2iip} there exists a protocol $S''$ of communication cost \asO{k+\log^2(n)} solving \Piip\ in \f0-error setting \wrt \Uii[2]\ with probability $\fr2{nk^2\log^2(n)}$.

Choose $t=2$, \theoref{t_hard_iip} implies that $S''$ can succeed only with probability \asO{\fr{k^2+\log^4(n)}{n^2}}, therefore $k\in\asOm{\fr{n^{\fr[]14}}{\sqrt{\log n}}}$, as required.}

\sect[s_open]{Conclusions and further work}
The protocol described in \sref{easy_qua} together with \theoref{t_main_low} imply \theoref{theo_main}.

It would be interesting to strengthen this result.
Is it possible to find a \e{functional} problem that requires exponentially more communication in \R\ than in \QI?
Raz~\cite{R99} constructs a \e{partial} function which is \e{complete}, in a natural and well-defined sense, for quantum one-way communication.
However, it is yet unclear what the classical complexity of Raz's function is.

It seems plausible that every \e{total} function with an efficient one-way quantum protocol admits an efficient classical protocol (maybe, even one-way).
Validity of this conjecture is a very important, well-known open problem.

What can be claimed about \R-complexity of communication problems with efficient quantum simultaneous protocols, either with or without shared entanglement?

As we have mentioned before, our \theoref{t_hard_iip} is a direct product statement and can be compared to other known direct product theorems, like that by Jain, Klauck and Nayak~\cite{JKN08_Dir} and earlier ones.
On the one hand, our statement is more rigorous, in the sense that it has plain ``$t$'' in the exponent of the guaranteed upper bound on the success probability of solving $t$ instances of the original problem, as opposed to ``\asOm{t}'' in the earlier works.
On the other hand, our theorem applies (or trivially generalizes) only to a restricted family of communication problems (those with structure similar to \D), as opposed to the result of \cite{JKN08_Dir}, which speaks about \e{any} communication problem.

Apparently, our technique can be applied to a wider class of communication problems, and, on the other hand, the approach taken in \cite{JKN08_Dir} can give a more rigorous statement in terms of $t$.
It would be interesting to analyze these two possibilities in order to give a more unified theory of direct product statements in communication complexity.

\subsection*{Acknowledgments}
This work has started from Richard Cleve's sharing with me his conjecture and Harry Buhrman's letting me know about the conjecture made by Serge Massar.
I would like to thank Alexander Razborov for finding a mistake in an early version, Ronald de Wolf for his help on improving the presentation, and an anonymous referee for many helpful comments.

\bib

\end{document}